\documentclass[aps,prd,a4paper,reprint,nofootinbib,superscriptaddress,floatfix,preprintnumbers]{revtex4-1}

\usepackage{hyperref}
\usepackage{amsmath}
\usepackage{amsfonts}
\usepackage{amssymb}
\usepackage{color}
\usepackage[utf8]{inputenc}
\usepackage{graphicx}
\usepackage{microtype}
\usepackage{siunitx}
\usepackage{soul}
\usepackage{comment}

\usepackage{lineno}

\newcommand{\Fermi}{\textit{Fermi}}
\newcommand{\Jf}{$J$}

\newcommand{\sv}{$\langle \sigma v \rangle$}

\def \aap  {A\&A}

\def \aj  {AJ}
\def \apj  {ApJ}
\def \apjs  {ApJS}
\def \apjl  {ApJL}

\def \jcap  {JCAP}
\def \prd {Phy. Rev. D}

\def \mnras {MNRAS}

\begin{document}

\begin{flushleft}
LAPTH-011/22
\end{flushleft}

\title{Investigating the effect of Milky Way dwarf spheroidal galaxies extension on dark matter searches with {\it Fermi}-LAT data}
\date{\today}

\author{Mattia Di Mauro }\email{dimauro.mattia@gmail.com}
\affiliation{INFN, Torino, Via Pietro Giuria 1, 10125}
\author{Martin Stref}\email{stref@lapth.cnrs.fr}
\affiliation{LAPTh, USMB, CNRS,  F-74940 Annecy, France}
\author{Francesca Calore}\email{calore@lapth.cnrs.fr}
\affiliation{LAPTh, USMB, CNRS,  F-74940 Annecy, France}

\begin{abstract}
Satellite galaxies of the Milky Way with high mass-to-light ratios and little baryon content, i.e.~dwarf spheroidal galaxies (dSphs), are among
the most promising targets to detect or constrain the nature of dark matter (DM) through its final annihilation products into high-energy photons.
Previously, the assumption that DM emission from dSphs is point-like has been used to set strong constraints on DM candidates using data from the \Fermi~Large Area Telescope (LAT). 
However, due to their high DM densities and proximity, dSphs actually have sufficient angular extension to be detected by the \Fermi-LAT.
Here, we perform a comprehensive analysis about the impact of accounting for angular extension in the search for gamma-ray DM signals towards known dSphs with \Fermi-LAT.
We show that, depending on the dSph under consideration, limits on the DM cross section can be weakened by up to a factor of 2--2.5, while the impact on the stacked, i.e.~combined, limits is at most 1.5--1.8 depending on the annihilation channel.
This result is of relevance when comparing dSphs limits to other multi-messenger DM constraints and for testing the DM interpretation of anomalous ``excesses''.

\end{abstract}

\maketitle

\section{Introduction}\label{intro}

Dark matter (DM) represents about the $85\%$ of matter in our
Universe \cite{Planck:2018vyg}, and yet its particle nature is a major puzzle for contemporary Physics.
This puzzle can be tackled from several corners.
Among them, indirect searches offer a unique way to probe different aspects of DM through a plethora of astroparticle observables, from cosmic surveys to fluxes of cosmic rays, see e.g.~\cite{Gaskins:2016cha,AlvesBatista:2021gzc}.

Traditionally, indirect searches look for signatures of cosmic photons and charged particles from GeV to TeV energies produced by DM annihilation or decay in space.
Indeed, DM, in the context of weakly interacting massive particles (WIMPs), is believed to annihilate or decay into standard model particles which are not stable but rapidly hadronise and/or decay producing fluxes of stable, observable, particles such as photons and cosmic rays (e.g.~positrons and antiprotons).
Signals of the DM production of cosmic particles are then searched for over the more abundant astrophysical background and foreground emissions.
Among the possible cosmic particles, photons have the advantage of direct propagation on Galactic scales and DM can be searched in the direction of specific astrophysical objects with predicted high DM density. 
Several DM searches have been performed in the last 14 years using gamma-ray data of the {\it Fermi} Large Area Telescope (LAT) in the direction of different astrophysical targets such as clusters of galaxies, Milky Way dwarf spheroidal galaxies (dSphs hereafter) irregular galaxies, the Milky Way halo, and the Galactic center, see e.g.~\cite{Fermi-LAT:2016afa} for an overview.
%
None of them has brought to a clear detection, and strong constraints on the DM particle properties have been set.

Perhaps the most promising targets so far used to identify 
(and/or constrain) the nature of DM are Milky Way dSphs, which are characterized by mass-to-light ratios in the range $10-1000$. Moreover, these objects are thought to have very little baryon content and possible astrophysical production of photons, i.e.~from pulsars \cite{Winter:2016wmy} (see~\cite{Strigari:2018utn} for a review).
DSphs have been targeted by several instruments, from radio wavelengths
to high-energy gamma rays, and have allowed us to set some of 
the strongest constraints in the annihilation
cross section vs mass plane for WIMP DM~\cite{Ackermann:2015zua,MAGIC2016,Fermi-LAT:2016uux,2019MNRAS.482.3480P,HoofEtAl2020,DiMauro:2021qcf}.
%
Nonetheless, in recent years, scientists have highlighted 
some limitations which weaken the robustness of the DM limits from dSphs.
First, statistical and systematic modeling uncertainties of the DM distribution 
in dSphs (e.g. contamination of foreground non-member stars and/or triaxiality) are especially important for ultra-faint objects, for which only hundreds of
member stars are detected~\cite{Bonnivard:2015xpq}.
Uncertainties related to departure from spherical symmetry and velocity anisotropy of the DM halo, 
as well as the effect of
contaminating foreground stars may significantly alter the predicted DM flux, and affect, in turn,
the limits by a factor of two to three~\cite{BonnivardEtAl2015a,UllioEtAl2016,SandersEtAl2016,BonnivardEtAl2016,HayashiEtAl2016,KlopEtAl2016,IchikawaEtAl2017}.
Secondly, systematic uncertainties associated with the modeling of the
astrophysical background at the dSph position can weaken the limits by a factor of a few, 
as in the case when assuming purely data-driven methods for background estimations~\cite{Mazziotta:2012ux,GeringerSamethEtAl2015,BoddyEtAl2018,Calore:2018sdx,Alvarez:2020cmw}.
Finally, the contamination from pulsars and millisecond pulsars may be larger than previously believed~\cite{2022arXiv220412054C}.

Most of the searches for a DM signals towards dSphs have been performed
by looking for excess of photon counts over the 
modeled astrophysical background matching a point-like DM signal
from the dSph direction (see, e.g., \cite{MAGIC2016,Fermi-LAT:2016uux,Calore:2018sdx,2019MNRAS.482.3480P,HoofEtAl2020,Alvarez:2020cmw,DiMauro:2021qcf}).
This was motivated by the fact that the size of the possible DM halos around dSphs is expected to be much smaller than the {\it Fermi}-LAT PSF below 1 GeV.
Refs.~\cite{Fermi-LAT:2013sme,Geringer-Sameth:2014qqa,Ackermann:2015zua} investigated the effect of the DM density profile extension in the analysis.
However, with several years of {\it Fermi}-LAT observations and the improved data selection of Pass 8 \cite{2018arXiv181011394B}, the size of extension of sources can be found, for relatively bright sources, to have values as low as $0.1^{\circ}-0.2^{\circ}$.

Source extension has been studied in the context
of searches for sub-halos in unidentified \Fermi~sources~\cite{Bertoni:2016hoh,Coronado-Blazquez:2019pny,2012JCAP...11..050Z,Biteau:2018tmv,Ciuca:2018vsz},
as well as included in the calculation of sensitivity predictions
with future gamma-ray instruments~\cite{Egorov:2018cip,Chou:2017wrw}.
In particular, Ref.~\cite{DiMauro:2020uos,2022PhRvD.105h3006C} explicitly showed that, in typical simulations of DM sub-halos, there is a correlation between the DM annihilation expected flux -- which is proportional to the so-called
\Jf-factor, the integral along the line of sight (l.o.s.) of the
DM density squared -- and the halo extension.
As noticed therein, this also naturally applies to dSphs, which 
are the more massive sub-halos, i.e.~with large \Jf-factors, and the smallest objects where star formation has been triggered.
Based on that and according to both semi-analytical and numerical simulations, DM sub-halos, and even more so dSphs,
can have an angular extension in the sky larger than the {\it Fermi}-LAT sensitivity for extended source detection~\cite{Coronado-Blazquez:2019pny,DiMauro:2020uos}.
Ref.~\cite{Gammaldi:2017mio,Gammaldi:2021zdm} studied the effect of the source extension on the geometrical factor for irregular galaxies and they found the constraints on a possible DM contribution by including the extension of the DM templates.
Therefore, the search for a DM signal in dSphs galaxies can be affected by the likely halo extension.

In this paper, we follow our previous work in~\cite{DiMauro:2020uos} and explore, for the first time, the impact of including halo extension on the DM limits using \Fermi-LAT data collected from the direction of known dSphs.
First, we calculate the expected effect using simulated data showing that the DM halo size indeed affects the upper limits on the annihilation cross section.
Then, we demonstrate that the effect found in simulations is confirmed with real data: Depending on the dSph extension and properly accounting for it can weaken the limits by up to a factor of 1.5 -- 1.8, depending on the annihilation channel. 
This result can impact the DM interpretation of the anomalous \Fermi-LAT Galactic center excess, see e.g.~\cite{2020ARNPS..70..455M} for a review.
In fact, the best-fit region for the DM mass and annihilation cross section that fit the GeV excess observations starts to be challenged by different, complementary, constraints on DM particle models set with other targets or other messengers.
If this tension is confirmed this may be a strong indication that the DM interpretation of this excess should be dismissed. 
As for dSphs, it has been shown that uncertainties of a factor of a few may worsen or alleviate this tension.
As we will show, the fact that including the dSphs extension weakens these limits by a factor up to 1.5 -- 1.8 may therefore be relevant to assess the tension between dSphs limits and the DM GeV excess best-fit region.
Finally, we also assess what is the impact of tri-axiality on the final limits when the full halo extension is considered, similarly to what was done in~\cite{KlopEtAl2016}.

The paper is organized as follows.
In Sec.~\ref{sec:model}, we present the set of dSphs
used in the present work and how we model the distribution of 
DM therein.
In Sec.~\ref{sec:datanalysis}, we describe 
the data selection and analysis technique, which 
we validate on mock data. 
Validation tests and results are presented in 
Sec.~\ref{sec:simdata}.
We finally illustrate our results in Sec.~\ref{sec:resdsphs},
and conclude in Sec.~\ref{sec:conclusions}.

\section{Dark matter density in Dwarf Spheroidal Galaxies}
\label{sec:model}

\subsection{Spherical templates}

Gamma-ray searches for DM annihilation in dSphs rely on the evaluation of the so-called $J$-factor 
\begin{eqnarray}
J(\Delta\Omega) = \int_{\Delta\Omega}\int_{\rm l.o.s.}\rho^2(l,\Omega)\,\mathrm{d}l\,\mathrm{d}\Omega\,,
\label{eq:geom}
\end{eqnarray}
where $l$ is the l.o.s.~coordinate, $\rho$ is the DM density, and $\Delta\Omega$ the solid angle over which integration is performed. In order to compute this $J$-factor, one needs to model the DM density inside dSphs. This is usually done adopting the Jeans equations and the observed dynamics of stars hosted by these systems. In this work, we rely on the mass modelling performed in two previous studies \cite{Alvarez:2020cmw,2019MNRAS.482.3480P}, and we consider a sample 
of 22 dSphs. 
These can be divided into two broad class: {\it classical} objects which contain hundreds to thousands of member stars, and {\it ultra-faint} objects which only possess tens of stars. Among our 22 dSphs, we have 8 classical dSphs and 14 ultra-faint dSphs. 

For classical dSphs we rely on Ref.~\cite{Alvarez:2020cmw}, 
which performed a Jeans analysis assuming spherical symmetry and steady-state for each object. The usual degeneracy between density and velocity anisotropy is lifted by considering higher-order Jeans equations~\cite{2017MNRAS.471.4541R}.
The DM density follows the {\tt coreNFW} functional form introduced in Ref.~\cite{Read:2015sta} 
\begin{eqnarray}
\rho_{\rm cNFW}(r) = f^n\,\rho_{\rm NFW} + \frac{n\,f^{n-1}(1-f^2)}{4\pi\,r^2\,r_c}\,M_{\rm NFW}\,,
\label{eq:coreNFW}
\end{eqnarray}
where $f = \tanh\left(r/r_c\right)$ and $r_c$ is the core radius. The quantities $\rho_{\rm NFW}$ and $M_{\rm NFW}$ refer to the density and mass of the well-known Navarro-Frenk-White (NFW) profile \cite{Navarro:1995iw} 
\begin{eqnarray}
\rho_{\rm NFW}(r) = \rho_s\,\frac{r_s}{r}\,\frac{1}{(1+r/r_s)^2}\,,
\label{eq:NFW}
\end{eqnarray}
where $\rho_s$ and $r_s$ are the scale density and scale radius, respectively. While NFW describes a system with a cuspy density profile, {\tt coreNFW} is flexible enough to describe both cored and cuspy systems.
The {\tt coreNFW} profile is further modified to account for tidal stripping as done in Ref.~\cite{Read:2018pft}
\begin{eqnarray}
\rho_{\rm cNFWt}(r) = \left\{
\begin{array}{lc}
\rho_{\rm cNFW}(r) & r\leqslant r_t \\
\rho_{\rm cNFW}(r_t)(r/r_t)^{-\delta} & r>r_t
\end{array}
\right.
\label{eq:coreNFWtides}
\end{eqnarray}
where $r_t$ is the tidal radius. This final form is referred to as {\tt coreNFWtides}. The DM profile in each dSph is thus characterized by 6 free parameters: $\rho_s$, $r_s$, $r_c$, $n$, $r_t$ and $\delta$. We use the Markov Chain Monte Carlo (MCMC) posterior chains provided by the authors of  \cite{Alvarez:2020cmw} to compute the median value of each parameter and the resulting $J_{05}=J(0.5^\circ)$\footnote{$J_{05}$ represents the value of the geometrical factor obtained by performing the integration of Eq.~(\ref{eq:geom}) as:
\begin{eqnarray}
J(\theta_{\rm max}) = 2\pi \int_0^{\theta_{\rm max}}\mathrm{d}\theta\,\sin\theta\int_{\rm l.o.s.}\rho^2(l,\Omega)\,\mathrm{d}l\,,
\end{eqnarray}
where $\theta_{\rm max}=0.5^{\circ}$.
}.

From the same posterior chains, one can compute the fully
data-driven probability distribution function (PDFs) of the \Jf-factor.
While stressing the relevance of
using these data-driven PDFs, Ref.~\cite{Alvarez:2020cmw} 
also checked that a log-normal fit provides a reasonable approximation to the \Jf-factor PDFs for classical dwarfs.
Since data-driven \Jf-factor PDFs
have not yet been derived for ultra-faint dSphs,  
for the sake of performing a global and consistent analysis over the sample of classical and ultra-faint dSphs, we have decided to adopt the log-normal approximation of the \Jf-factor PDF for both classical and ultra-faint dSphs.

We quote the $J_{05}$ and corresponding one-standard-deviation uncertainties from the log-normal fit in the 8 top rows of Tab.~\ref{tab:jfactors}. We also report the total geometrical factor $J_{\rm{tot}}$ which is integrated up to $5\times r_{\rm t}$ to account for the DM located beyond $r_{\rm t}$, see Eq.~(\ref{eq:coreNFWtides}), and the corresponding uncertainty.

For ultra-faint dSphs, we refer to Ref.~\cite{2019MNRAS.482.3480P}.
There, the authors performed a Jeans analysis on a large number of dSphs. Equilibrium and spherical symmetry are also assumed, while the anisotropy is a free constant parameter. The DM profile follows the NFW profile in Eq.~(\ref{eq:NFW}) thus it is characterized by 2 free parameters $\rho_s$ and $r_s$. The profile is sharply truncated at the tidal radius $r_{\rm t}$. Unlike classical dSphs, $r_{\rm t}$ for the ultra-faint dSphs is not directly fitted but instead computed using the formula $r_{\rm t}=\left[M_{\rm sub}(r_{\rm t})/(2-\mathrm{d}\ln M_{\rm host}/\mathrm{d}\ln R)\,M_{\rm host}\right]^{1/3}\,R$ where $R$ is the radial position of the dSph within the Galaxy and $M_{\rm host}$ is the total mass within that radius. The tidal radius implicitly depends on $\rho_{\rm s}$, $r_{\rm s}$ and the distance $D$ from the dSph which is a nuisance parameter of the analysis. 
We use the publicly available MCMC posterior chains to compute the median value of these parameters. We exclude a number of objects from the analysis of \cite{2019MNRAS.482.3480P}: We remove dSphs which have an unresolved or only partially resolved l.o.s.~velocity dispersion. Since we are interested in the impact of extended DM templates, we also remove objects that are not satellites of the Milky Way and are too far away to show any significant extension. In fact, we discard objects with a distance $>300\,\rm kpc$. 
We are thus left with 14 dSphs which are listed along with their $J_{05}$ in the 14 bottom rows of Tab.~\ref{tab:jfactors}. We also report $J_{\rm tot}$ which is integrated up to the tidal radius as a sharp truncation of the profile at $r_{\rm t}$ is assumed for the ultra-faints dSphs. Uncertainties on both $J_{05}$ and $J_{\rm tot}$ are obtained by fitting a lognormal PDF through the corresponding distribution.

Before proceeding to the detailed analysis, we can already single out targets which can be significantly extended. We do this by computing the angle $\theta_{68}$ which contains 68$\%$ of the total $J$-factor
\begin{eqnarray}
J(\theta_{68}) \equiv 0.68\times J_{\rm tot}\,.
\end{eqnarray}
This angle is computed for each dSph template and the result is shown in the right column of Tab.~\ref{tab:jfactors}. Very roughly, we expect limits set from objects with $\theta_{68}\gtrsim 0.5^\circ$ to be impacted by the use of an extended template in place of a point-like one. In particular, 
Sculptor among the classical dSphs and Ursa Major II among the ultra-faint ones show the largest extensions, 0.65$^\circ$ and 0.74$^\circ$ respectively.
Note that for most targets $\theta_{68}$ is much smaller than the physical angular extension $\theta_{\rm tot}$ set by the tidal radius\footnote{This is not strictly the case for the {\tt coreNFWtides} template which has a density that goes smoothly to zero at infinity.}. 
For the usual thermal relic cross section, the DM density near $r_{\rm t}$ is much too low for the annihilation to be detectable so $\theta_{68}$ is a better proxy for the detectable extension of an object. Nevertheless, we provide $\theta_{\rm tot}$ and the distance $D$ to the source in Tab.~\ref{tab:jfactors}.
We note that $\theta_{\rm tot}$ is lower for classical dSphs than for ultra-faint dSphs, which can be traced back to lower values of $r_{\rm t}$. We recall that for classical dSphs $r_{\rm t}$ is simply a fitting parameter, which potentially underestimates the true tidal radius. This has no consequence on our analysis since $\theta_{68}$ is a more relevant parameter.

As a concluding remark for this Section, we would like to point out that the angular size (or $\theta_{68\%}$) is an effective parameter which depends on the fundamental dSph parameters, namely the distance and the DM spatial profile.  By virtue of the definition of $\theta_{68\%}$, cuspier profiles produce a smaller $\theta_{68\%}$ be since more flux is contained in a smaller angular size. So, if the DM profile’s parameters (and parameterizations) of the dSphs significantly differ from the ones used here, they will predict a different $\theta_{68\%}$ and yield a different impact of the extension on the single dSph DM limits.
We stress, that, for the present work we have made use of the latest and, presumably, most robust analyses 
for the determination of the mass distribution in dSphs. 
For the sake of completeness, we have reported the median values of the DM density 
parameters for each dSph in Appendix~\ref{app:profile_parameters}.

\subsection{Non-spherical templates}
\label{sec:triaxial_template}
There is observational evidence for non-sphericity of the luminous halo of several dSphs \cite{McConnachie2012,Walker2013,SanchezJanssenEtAl2016}.
Furthermore, cold DM-only cosmological simulations show that the DM profile of satellite galaxies are in general not spherical but instead mildly triaxial \cite{KuhlenEtAl2007,VeraCiroEtAl2014}, although baryonic feedback effects can make these halos more spherical \cite{AbadiEtAl2010,ZempEtAl2012}.
Since departures from spherical symmetry in the dSph DM profiles are known to be an important source of uncertainty when setting constraints on the annihilation cross section \cite{BonnivardEtAl2015a,SandersEtAl2016,HayashiEtAl2016,KlopEtAl2016}, we also consider a triaxial template in our analysis.
Such a template is created by simply replacing the spherical radius $r$ in Eq.~(\ref{eq:coreNFWtides}) by the ellipsoidal radius:
\begin{equation}
    r \rightarrow \sqrt{\left(\frac{x}{a}\right)^2
    +\left(\frac{y}{b}\right)^2
    +\left(\frac{z}{c}\right)^2}
\end{equation}
where $a$, $b$ and $c$ are the axis parameters with $a\geqslant b\geqslant c$ and $abc=1$. 
We fix the axis ratios to $b/a = 0.8$ and $c/a = 0.6$ which are values close to the ones found in simulations \cite{KuhlenEtAl2007,VeraCiroEtAl2014} and are also used in the triaxial analysis performed by Ref.~\cite{BonnivardEtAl2015a}. 

We keep the values of the profile parameters ($\rho_{\rm s}$, $r_{\rm s}$, etc.) obtained from the spherical Jeans analysis. This is not entirely consistent as one should instead re-do the Jeans analysis on the data starting from the triaxial ansatz instead of the spherical one. Our goal here however is not to provide the most realistic description but rather to gauge the general impact of triaxiality for an extended target. 
In the following, we consider three extreme configurations corresponding to the l.o.s.~being aligned with either the major, second or minor axis.

\begin{table*}[]
    \centering
    \begin{tabular}{l|c|c|c|c|c}
           & ${\rm log_{10}}(J_{05})$ & ${\rm log_{10}}(J_{\rm tot})$ & $\theta_{68}$ & $\theta_{\rm tot}$ & {$D$} \\
          & [$\rm GeV^2/cm^5/sr$] & [$\rm GeV^2/cm^5/sr$]  & [$^\circ$] & [$^\circ$] & [$\rm kpc$] \\  
 \hline
 Ursa Minor & $18.31\pm0.08$ & 18.55$\pm0.05$ & 0.59 & 0.84 & 76 \\
 Draco & $18.64\pm0.04$ & 18.73$\pm0.03$ & 0.35 & 0.84 & 76 \\
 Sculptor & $18.39\pm0.05$ & 18.67$\pm$0.09 & 0.65 & 1.02 & 86 \\
 Sextans & $18.07\pm0.08$ & 18.15$\pm0.06$ & 0.35 & 0.84 & 86 \\
 Leo I & $17.50\pm0.06$ & 17.52$\pm0.06$ & 0.12 & 0.31 & 254 \\
 Leo II & $17.51\pm0.05$ & 17.51$\pm0.05$ & 0.07 & 0.13 & 233 \\
 Carina & $17.92\pm0.07$ & 18.01$\pm0.11$ & 0.36 & 0.88 & 105 \\
 Fornax & $17.76\pm0.05$ & 18.00$\pm0.07$ & 0.59 & 0.94 & 138 \\
 \hline
 Aquarius II & $18.26\pm0.62$ & 18.30$\pm0.67$ & 0.19 & 5.54 & 108\\
 Bootes I & $18.17\pm0.30$ & 18.34$\pm0.41$ & 0.52 & 5.41 & 66 \\
 Canes Ven.\ I & $17.35\pm0.16$ & 17.39$\pm0.21$ & 0.17 & 5.90 & 210 \\
 Canes Ven.\ II & $17.84\pm0.53$ & 17.92$\pm0.60$ & 0.25 & 7.05 & 160\\
 Carina II & $18.22\pm0.58$ & 18.34$\pm0.66$ & 0.38 & 3.21 & 37\\
 Coma Beren. & $19.01\pm0.38$ & 19.21$\pm0.55$ & 0.58 & 6.59 & 42\\
 Hercules & $17.30\pm0.54$ & 17.32$\pm0.57$ & 0.11 & 3.19 & 132\\
 Horologium I & $18.68\pm1.02$ & 18.70$\pm1.06$ & 0.13 & 4.94 & 87\\
 Reticulum II & $18.92\pm0.41$ & 19.09$\pm0.62$ & 0.51 & 4.70 & 32\\
 Segue 1 & $18.96\pm0.71$ & 19.00$\pm0.77$ & 0.17 & 2.84 & 23\\
 Tucana II & $18.83\pm0.56$ & 19.03$\pm 0.63$ & 0.57 & 6.76 & 57\\
 Ursa Major I & $18.22\pm0.29$ & 18.28$\pm0.34$ & 0.24 & 5.85 & 97\\
 Ursa Major II & $19.46\pm0.41$ & 19.71$\pm0.53$ & 0.74 & 8.78 & 35 \\
 Willman 1 & $19.52\pm0.55$ & 19.59$\pm0.70$ & 0.24 & 5.68 & 38
    \end{tabular}
\caption{Sample of dSphs used in this study with their associated $J_{05}$, $J_{\rm tot}$, $\theta_{68}$, $\theta_{\rm tot}$ and distance $D$. DSphs in the top rows are taken from \cite{Alvarez:2020cmw}, {\it classical} dSphs, while dSphs in the bottom rows are taken from Ref.~\cite{2019MNRAS.482.3480P}, {\it ultra-faint} dSphs.
    }
    \label{tab:jfactors}
\end{table*}


\section{Data selection and analysis technique}
\label{sec:datanalysis}

\subsection{Data selection}
\label{sec:data}

We perform our analysis with twelve years\footnote{Mission Elapsed Time (MET): 239557417 s $-$ 618050000 s} of Pass 8 \Fermi-LAT data with the P8R3 processing.
We select SOURCEVETO class events\footnote{SOURCEVETO is an event class recently created by the {\it Fermi}-LAT Collaboration to maximize the acceptance while minimizing, at the same time, the irreducible cosmic-ray background contamination. In fact, SOURCEVETO class has the same contamination level of P8R2\_ULTRACLEANVETO\_V6 class while maintaining the acceptance of P8R2\_CLEAN\_V6 class.}, passing the basic quality filter cuts\footnote{DATA\_QUAL$>$0 \&\& LAT\_CONFIG==1}, and their corresponding P8R3\_SOURCEVETO\_V2 response functions.
We choose energies between 0.5 to 1000 GeV and apply a cut to zenith angles $<100^\circ$ in order to exclude contamination from the limb of the Earth.
We decide to start our analysis from 0.5 GeV because we want to investigate the effect of the extension of dSphs. 
Including data with energies $<0.5$ GeV, where the PSF is much larger, would not improve the sensitivity of our results.
In fact the angular resolution below 500 MeV is typically larger than $1^{\circ}$ while above 1 GeV could be as low as $0.1^{\circ}$.
For each target in our analysis, we select a $14\times14$ deg$^2$ region of interest (ROI) centered at the dSphs position and choose pixel size of $0.08$ deg. 
We only consider spherical templates in this section. Uncertainties associated to triaxiality will be discussed in Sec.~\ref{sec:sys_asymm}.
%


\subsection{Analysis technique}
\label{sec:analysis}
The DM search in our sample of dSphs follows the analysis performed in the past by the {\it Fermi}-LAT Collaboration on these sources (see, e.g., \cite{Ackermann:2015zua}) or more recently in the direction of Andromeda and Triangulum galaxies \cite{DiMauro:2019frs}.
We provide a general overview and we refer to Refs.~\cite{Ackermann:2015zua,DiMauro:2019frs} for a complete description of the analysis technique.
We use the public \textit{Fermipy} package (version 0.19.0) to perform a binned analysis with eight bins per energy decade. 
\textit{Fermipy} is a python wrapper of the official {\tt Fermitools}, for which we use version 1.3.8.

In each of the 22 dSph ROIs, which we analyze independently\footnote{See Ref.~\cite{Calore:2018sdx} for some limitations related to independent ROI fits.}, we model the total gamma-ray emission as the sum 
of {\it background} plus {\it signal} events.
The astrophysical \emph{background} model is made up by: (1) Sources as reported in the 10-year Source Catalog (4FGL-DR2)\footnote{\url{https://arxiv.org/pdf/2005.11208.pdf}} 
including sources located at most $2^{\circ}$ outside our ROI,
(2) the latest released interstellar emission model (IEM), namely {\tt gll\_iem\_v07.fits}\footnote{A complete discussion about this new IEM can be found at \url{https://fermi.gsfc.nasa.gov/ssc/data/analysis/software/aux/4fgl/Galactic_Diffuse_Emission_Model_for_the_4FGL_Catalog_Analysis.pdf}}, and (3) its corresponding isotropic template {\tt iso\_P8R3\_SOURCEVETO\_V3\_v1.txt}.
The {\it signal} we look for is an additional source at each dSph position. To model the additional source term, we consider two scenarios: 
(a) the point-like source case (\texttt{PS} hereafter), where
the new source has no extension, 
and (b) the extended case (\texttt{Ext} hereafter), where 
the additional source spatial distribution is fully included in the fit by making use of the extended templates described in Sec.~\ref{sec:model}.

We perform the following analysis' steps:
\begin{itemize}
\item[1.] {\it Optimization of background model in dSPhs ROIs}. A baseline fit is performed on each ROI including sources in the 4FGL-DR2, IEM and isotropic template.
A refinement of the model is run by relocalizing all point-like sources in the model. We check that the new positions are compatible with the ones reported in the 4FGL-DR2 catalog. Then, we search for new point-like sources with a Test Statistic\footnote{The Test Statistic ($TS$) is defined as twice the difference in maximum log-likelihood between the null hypothesis (i.e., no source present) and the test hypothesis: $TS = 2 ( \log\mathcal{L}_{\rm test} -
  \log\mathcal{L}_{\rm null} )$~\cite{1996ApJ...461..396M}.} ($TS$) $TS >  25$ and distance at least $0.5^{\circ}$ from the center of the ROI. A final fit is then performed, where all the spectral energy distribution (SED) parameters of the sources, normalization and spectral index of the IEM and normalization of the isotropic component are free to vary. With this first step we thus have a background model that represents properly the gamma-ray emission in the ROI. In fact, in all the ROIs considered the residuals found by performing a $TS$ map with the background-only model are at most at the level of $\sqrt{TS}\sim 2-3$. These remaining residuals, if located close to the region of interest, could generate a small signal for the detection of the dSphs.  
\item[2.] {\it SED of additional source at dSph position}.
The additional source associated with DM emission at the position of each dSph is added in the center of the ROI either as point-like source (\texttt{PS} case) or as an extended source (\texttt{Ext} case). 
A fit with the background plus signal model is then performed for the two scenarios in each dSph ROI. 
The SED for the additional sources at the dSphs positions is calculated by performing a fit energy bin by energy bin. Specifically, the SED run gives for each energy bin the value of the likelihood as a function of the photon energy flux, $d\Phi_{\rm dSph}/dE$. With the SED information we can thus test every possible spectrum for the source of interest, including the DM one.
\item[3.] {\it Conversion from source energy flux to DM parameter space}. 
The flux of gamma rays produced from DM particles annihilation is: 
\begin{equation}
    \frac{d\Phi_{\rm DM}}{dE} = \frac{1}{4\pi}  \frac{\langle\sigma v \rangle}{2 M_{\rm{DM}}^2} J \times     \sum_f Br_f   \left( \frac{dN_\gamma}{dE} \right)_f
    \label{eq:fluxDM}
\end{equation}
where $M_{\rm{DM}}$ is the DM mass, $\langle \sigma v \rangle$ defines the annihilation cross section times the relative velocity, averaged over the Galactic velocity distribution function and $J$ is the geometrical factor.
$(dN_\gamma/d E)_f$ is the gamma-ray spectrum from DM annihilation for a specific annihilation channel labeled as $f$ and $Br_f$ is its branching ratio. 
We take $(dN_\gamma/d E)_f$ from Ref.~\cite{2008JCAP...11..003J} as implemented in the {\tt fermitools}\footnote{See the following page for a complete description of the DM model \url{https://fermi.gsfc.nasa.gov/ssc/data/analysis/scitools/source_models.html}.}.
%
We comment about the choice of \Jf-factor parameters and the impact on final constraints in Sec.~\ref{sec:simdata}.
We consider two DM annihilation channels with a branching ratio equal to 1, $b$-quarks and $\tau$-leptons pair annihilation, which correspond to the
most extreme behaviors of the DM SED and should bracket the DM spectral uncertainties.  
%
We use the SED information obtained in step (2) to calculate, for every annihilation channel, the likelihood as a function of annihilation cross section and DM mass values. 
We perform this analysis for each individual source in our sample.
For a given DM annihilation channel and mass the theoretical DM SED shape is fixed and for different values of the annihilation cross section ($\langle \sigma v \rangle$) we extract the corresponding likelihood from the SED data. 
\item[4.] {\it Extracting the $TS$ for the detection of DM or upper limits for $\langle \sigma v \rangle$}. 
For each individual dSph, we therefore obtain the likelihood as a function of DM mass and annihilation cross section.
The DM detection $TS$ is found by finding the maximum of the likelihood in the $\langle \sigma v \rangle$ and DM mass ($M_{\rm{DM}}$) space and comparing it with the likelihood of the null hypothesis, i.e. the one of the optimized ROI fit without DM emission. The upper limits of $\langle \sigma v \rangle$ are instead calculated in the following way. For a fixed DM mass, we take the likelihood profile as a function of $\langle \sigma v \rangle$, $\mathcal{L}(\langle \sigma v \rangle)$. We then can calculate the upper limits for $\langle \sigma v \rangle$ by finding the minimum of $\mathcal{L}(\langle \sigma v \rangle)$ and calculating the $\langle \sigma v \rangle$ that worsens the best-fit likelihood value by $\Delta \mathcal{L} = 2.71/2$\footnote{The fluxes for DM are taken to be non negative in our analysis. Therefore, the $\Delta \chi^2$ or equivalently the $2\Delta \mathcal{L}$ between the test and null hypothesis associated to the 95$\%$ CL is 2.71.}, which is associated with the one-sided $95\%$ CL upper limits. 
In finding the $TS$ or the upper limits for $\langle \sigma v \rangle$, we add to the Poissonian term of the likelihood a factor that takes into account the uncertainty on the \Jf-factor, assuming a log-normal distribution of this quantity~\cite{Ackermann:2015zua}:
\begin{widetext}
\begin{equation}
\label{eq:Lsigma}
    \mathcal{L}_i\left(J_i|J_{\rm{dyn},i},\sigma_i\right)= \frac{1}{\mathrm{log}(10)J_{\rm{dyn},i} \sqrt{2\pi}\sigma_i}   
    \times \exp{\left[- \left( \frac{\mathrm{log}_{10}(J_i) - \mathrm{log}_{10}(J_{\rm{dyn},i})}{\sqrt{2}\sigma_i} \right)^2\right]} \, ,
\end{equation}
\end{widetext}
where $J_{\rm{dyn},i}$ is the best fit for the dynamical geometrical factor for the $i$-th dwarf while $\sigma_i$ is the error in $\mathrm{log}_{10}(J_{\rm{dyn},i})$ space. Instead $J_i$ is the value of the geometrical factor for which the likelihood is calculated. 

According to standard practice, we profile over the \Jf-factor uncertainty.
This term of $\mathcal{L}$ disfavors values of $J_i$ much different from the observed one weighting it for the corresponding error.
We notice that the \Jf-factor parameters for the \texttt{PS} case and the
\texttt{Ext} do not need to match, 
and indeed we expect them to differ if the source has an extension larger than $0.5^\circ$.
For each dSph, the parameters of interest are: $J_{\rm{dyn}}^\texttt{PS}$, $\sigma^\texttt{PS}$, $J_{\rm{dyn}}^\texttt{Ext}$, $\sigma^\texttt{Ext}$. 
We discuss the choice of the parameters' values in Sec.~\ref{sec:simdata}.

Finally, we combine the results obtained by summing all dSphs' likelihoods. The same procedure as the single dSph case is then applied to derive the {\it stacked} $TS$ and upper limits on the annihilation cross section.
\end{itemize}

\subsection{Mock data generation}
\label{sec:mock}
For the sake of quantifying the impact of extension, we first run the full analysis chain on a set of mock data.
We build simulated data based on the optimized background emission model (1) in each dSph ROI, and we create multiple data sets by randomizing the counts in each pixel following the Poisson statistics. 

We then run the full analysis pipeline (1 -- 4) on this mock data set 
to quantify what is the sensitivity to a putative DM signal at the dSphs' positions. 

\section{Validity tests on simulated data}
\label{sec:simdata}
We here present the results of the validity tests 
performed on simulated data, generated according the 
procedure described in Sec.~\ref{sec:datanalysis}.
We follow the analysis' steps sketched in Sec.~\ref{sec:analysis}
for both \texttt{PS} and \texttt{Ext} scenarios.
The goals here are to 
assess how the upper limits on $\langle \sigma v \rangle$ 
change when varying the
\Jf-factor parameters in the likelihood (Eq.~\ref{eq:Lsigma}), or when  assuming an extended template for the DM flux instead of a PS one.
To isolate these effects, we consider a case
where we have the background model perfectly under control. 
%
%

We compute the 95\% C.L.~upper limits on \sv~separately for the \texttt{PS}
and \texttt{Ext} cases. 
In order to disentangle different effects, we consider the following three cases:
\begin{itemize}
    \item \texttt{Case 1}: We assume that the geometrical factor average value and error for the \texttt{Ext} and \texttt{PS} cases are the same: 
    $J_{\rm{dyn}}^\texttt{PS} = J_{\rm{dyn}}^\texttt{Ext} = J_{05}$ , and $\sigma^\texttt{PS} = \sigma^\texttt{Ext} = \sigma_{J_{05}}$, with values as in Tab.~\ref{tab:jfactors}. We stress that this choice of parameters is nonphysical since the \texttt{Ext} and \texttt{PS} \Jf-factors must have a different normalization by construction. Nonetheless, this case allows us to isolate the impact of the use of an extended template in the analysis.
    \item \texttt{Case 2}: We assume a different \Jf-factor average value for the \texttt{Ext} and \texttt{PS} case, while we keep the same error for the two cases:  $J_{\rm{dyn}}^\texttt{PS} =J_{05}$, $ J_{\rm{dyn}}^\texttt{Ext} = J_{\rm tot}$, $\sigma^\texttt{PS} = \sigma^\texttt{Ext} = \sigma_{J_{05}}$. Parameters' values as in Tab.~\ref{tab:jfactors}. 
    This case is performed to test how the results change taking into account both the extended spatial dSph template, as well as the corresponding different \Jf-factor normalization for the \texttt{Ext} and \texttt{PS} models.
    \item \texttt{Case 3} (baseline): Both the \Jf-factor average and error are different for the \texttt{Ext} and \texttt{PS} cases: 
    $J_{\rm{dyn}}^\texttt{PS} =J_{05}$, $ J_{\rm{dyn}}^\texttt{Ext} = J_{\rm tot}$, $\sigma^\texttt{PS} = \sigma_{J_{05}}$, and $\sigma^\texttt{Ext} = \sigma_{J_{\rm tot}}$. Parameters' values are as in Tab.~\ref{tab:jfactors}. 
    This is the most self-consistent choice of parameters.
    Indeed, for the \texttt{PS} case, this choice matches the one of previous works \cite{Ackermann:2015zua,2019MNRAS.482.3480P}, and can be motivated by the LAT angular resolution. 
    For the \texttt{Ext} case, instead, since the spatial template corresponds to the full DM halo extension, then the most self-consistent choice is to normalize this model with $J_{\rm tot}$. 
\end{itemize}

\begin{figure*}
\includegraphics[width=0.49\textwidth]{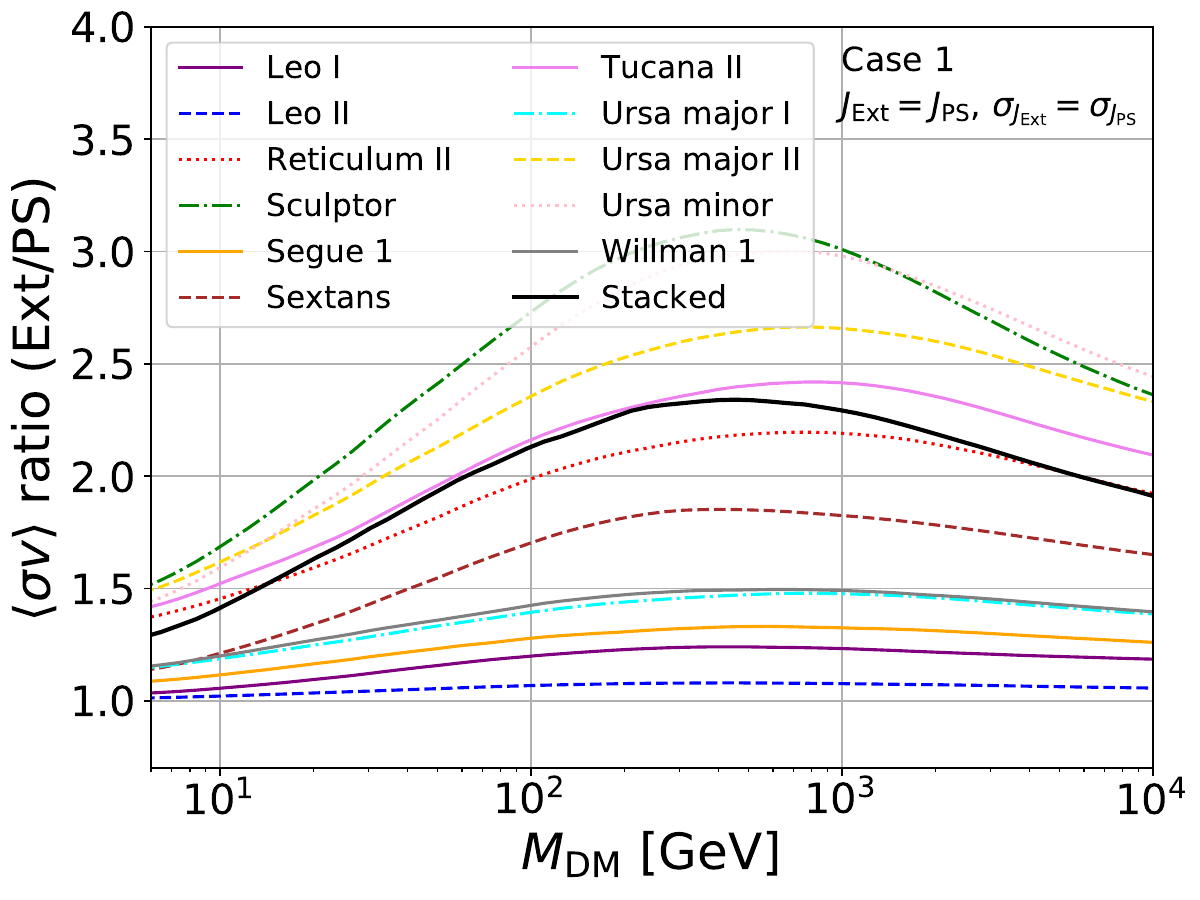}
\includegraphics[width=0.49\textwidth]{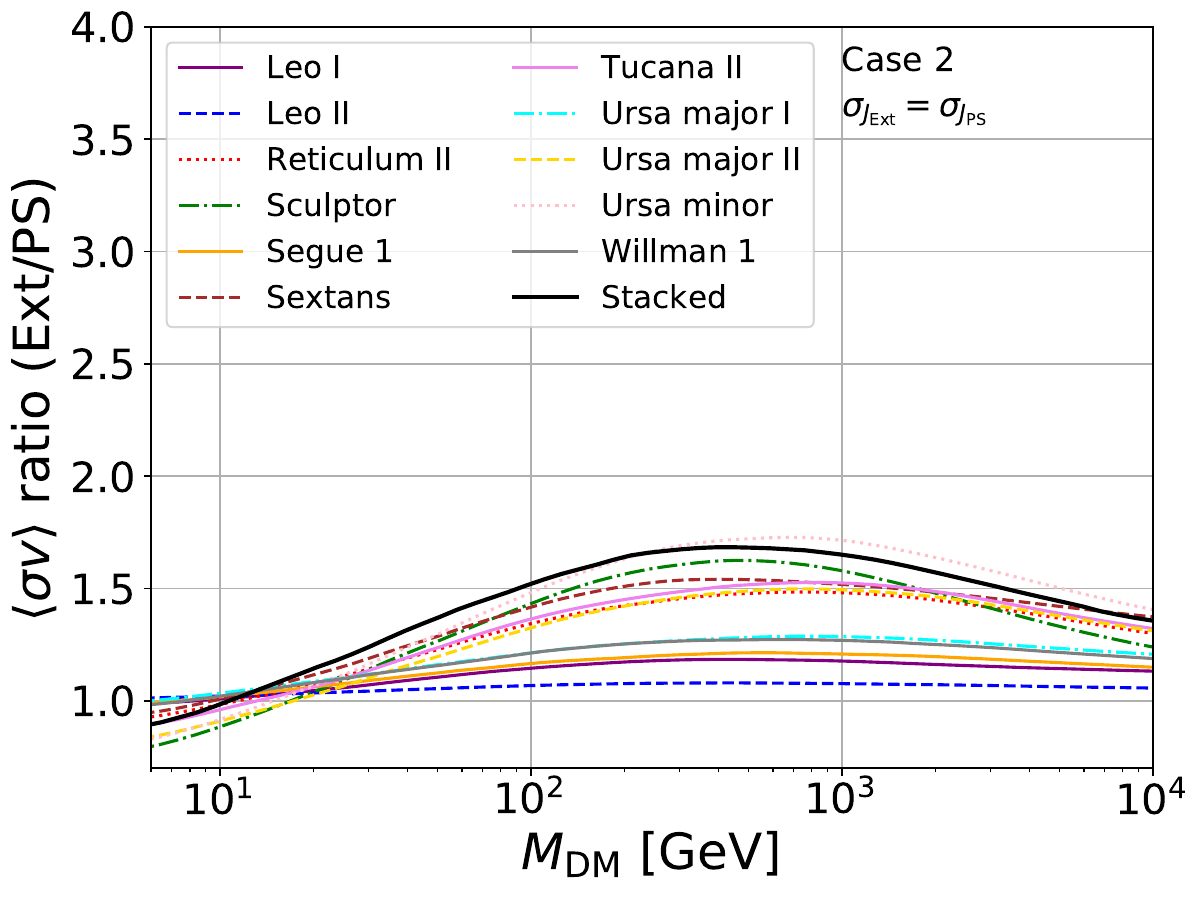}
\includegraphics[width=0.49\textwidth]{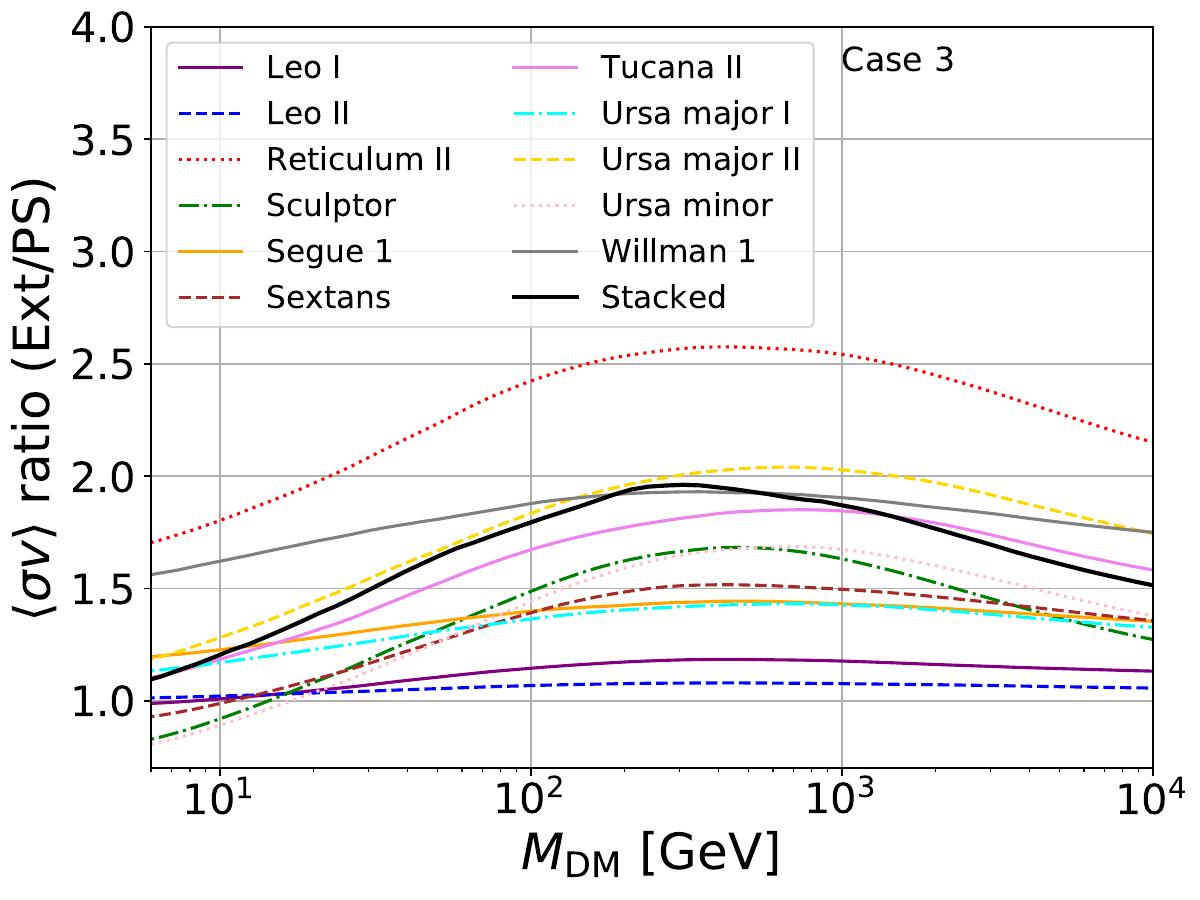}
\includegraphics[width=0.49\textwidth]{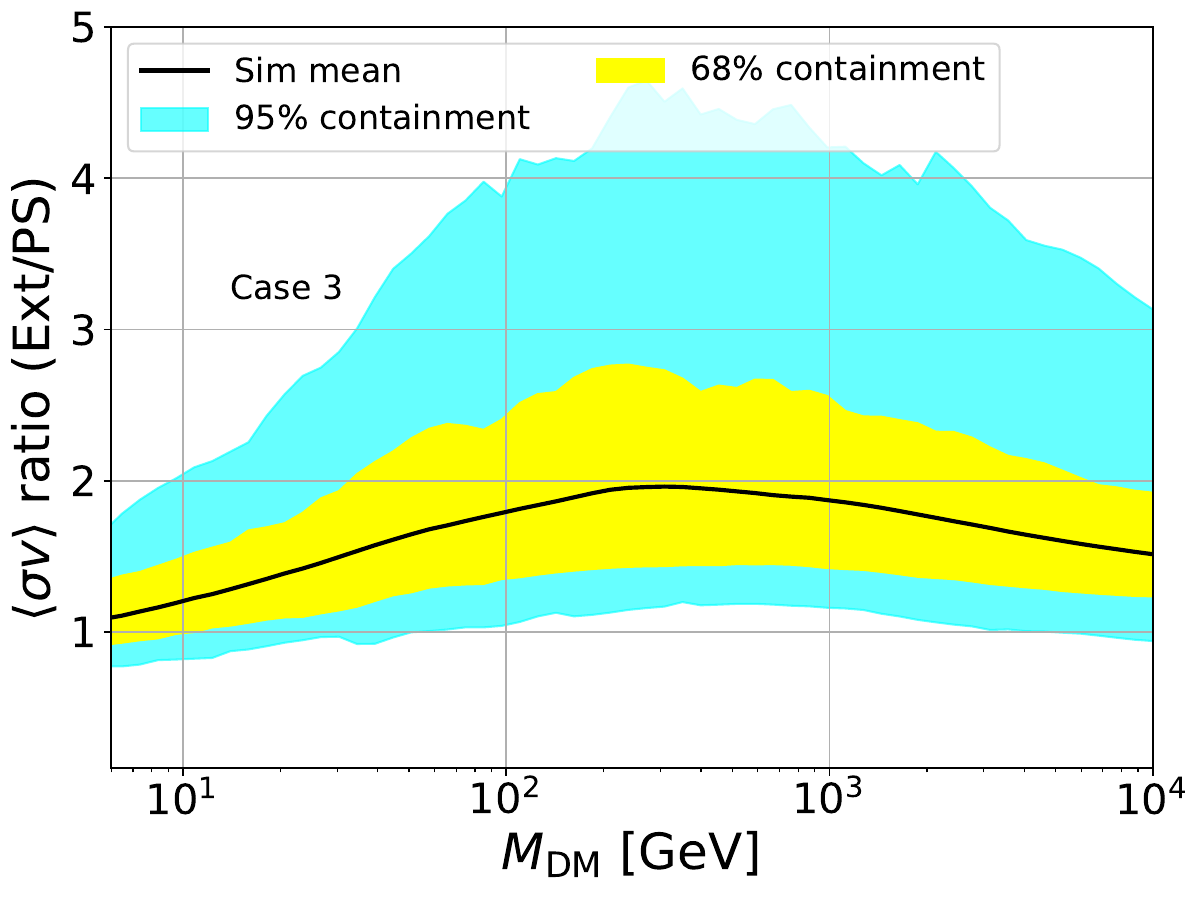}
\caption{{\it Simulated data}: Ratio of \sv~limits \texttt{Ext}/ \texttt{PS} for different choices of \Jf-factor parameters' values, see description in Sec.~\ref{sec:simdata}. Top left: For one half of the dSphs sample and total stacked result (black solid line) we show \texttt{Case 1} in the top left panel,  \texttt{Case 2} in the top right one,  and \texttt{Case 3} in the bottom left one. The stacked result for \texttt{Case 3}, together with the corresponding 68\% and 95\% C.L.~bands is displayed in the bottom right panel.
}
\label{fig:sim_ratio}
\end{figure*}

\begin{figure}
\includegraphics[width=0.49\textwidth]{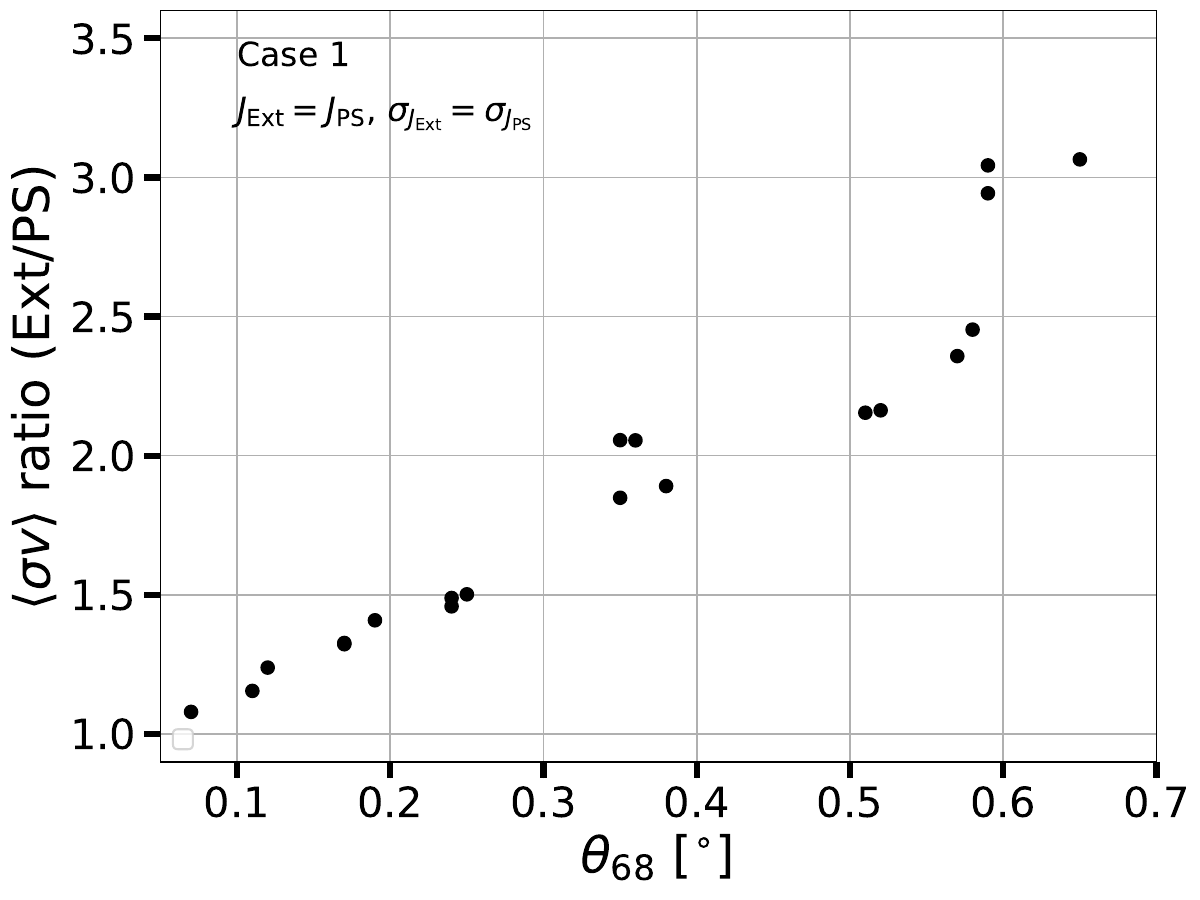}
\caption{Ratio of \sv~limits for the \texttt{Ext}/\texttt{PS} cases as a function of the parameter $\theta_{68}$ as reported in Tab.~\ref{tab:jfactors}. We show the results obtained for the {\tt Case 1}.}
\label{fig:correlation}
\end{figure}

We show the results obtained in the three cases in Fig.~\ref{fig:sim_ratio} for the parameter \sv~ratio \texttt{Ext}/\texttt{PS} for half of our {\it simulated} dSphs.
Similar conclusions are derived by using the other half of the dSphs sample.
In \texttt{Case 1} (top left), used to isolate the effect of the extended spatial template, we find that the ratio between the cross section in the \texttt{Ext} and \texttt{PS} models is always larger than one. 
This implies that the limits in the \texttt{Ext} case are always weaker than the ones in the \texttt{PS} case.
For most dSphs, the ratio is between 1.0 and 1.3 at low DM masses, and increases up to 2.0 -- 3.0 for masses between 100 GeV -- 1 TeV, as for e.g.~Sculptor, Ursa Minor, Fornax and Ursa Major II.\footnote{As highlighted above, the effect of the extension on the single 
dSphs depends on $\theta_{68\%}$, which ultimately depends on the DM profile of the dSphs. So any ranking of dSphs mentioned here has to be understood within the dSph mass modeling adopted in this work.}
This is explained by the fact that these sources are the most extended ones, see parameter $\theta_{68}$ in Tab.~\ref{tab:jfactors}. The ranking of the sources in the ratio of \sv~for \texttt{Ext} and \texttt{PS} models is following exactly the ranking of the parameter $\theta_{68}$.
We show this result in Fig.~\ref{fig:correlation}, where we report the ratios of the upper limits for \sv~obtained for the \texttt{Ext} and \texttt{PS} cases as a function of the parameter $\theta_{68}$ (see Tab.~\ref{tab:jfactors}).
A similar mass dependence is found in the stacked analysis shown with a black solid line in Fig.~\ref{fig:sim_ratio}. In this case the ratio reaches a maximum of about 2.3 at 400 GeV.

The mass dependence of the ratio can be understood as follows: The {\it Fermi}-LAT PSF at low energy is much larger than the one at high energy.
Moreover, DM particles with mass below 10 GeV have spectra that peak at low energy where {\it Fermi}-LAT has a poor resolution.
Therefore, in this mass regime the point-like source or the extended templates pick up roughly the same flux and, as a consequence, the ratio of the upper limits is expected to be about one (or in the other cases to trace the difference between $J_{\rm tot}$ and $J_{05}$).  In particular, the sources for which this ratio is the smallest, very close to one, are the dSphs with the smallest $\theta_{68}$.
At masses of a few hundreds of GeV the DM energy spectra peak at a few tens of GeV where the {\it Fermi}-LAT angular resolution is much better. In this case the point-like source template absorbs less photons than the extended template and in turn the value of $\langle \sigma v \rangle$ for the \texttt{PS} case are smaller than that of the \texttt{Ext} case and the ratio becomes larger than 1.
This effect is typically larger for dSphs with a more extended DM template.

In the \texttt{Case 2} (top right), where we use the same errors for the geometrical factor but different \Jf-factor average values, the ratio of \sv~is driven by a combination of two effects: The different extended template and the different values of $J$. For DM masses larger than 15 GeV, the limits in the \texttt{Ext} case are always weaker than the ones in the \texttt{PS} case, confirming that the strengthening of the limits at low masses for some dSphs is driven by the $J_{\rm tot}/J_{05}$ ratio.
At low masses, $M_{\rm{DM}}<15$ GeV, however, the ratio of the cross sections is systematically smaller than 1 for most of the sources. In fact, at such low masses the flux from DM is peaked at very low energy where the \texttt{Ext} and \texttt{PS} models convolved with the very poor PSF appear to have the same extension. Therefore, the ratio of \sv~is driven mainly by $J_{\rm{tot}}/J_{\rm{05}}$.
The sources with the largest $\theta_{68}$ are also the ones with the smallest \texttt{Ext}/\texttt{PS} for these values of the DM mass.
In this regime the limits on \sv~are weaker in the \texttt{PS} case with respect to the \texttt{Ext} approximation.  
Similarly to what was obtained in \texttt{Case 1}, the peak of the \sv~ratio is at masses of around 100 -- 1000 GeV and takes maximum values of about 1.3-1.7 for the dSphs that are the most extended.

\begin{figure}
\includegraphics[width=0.49\textwidth]{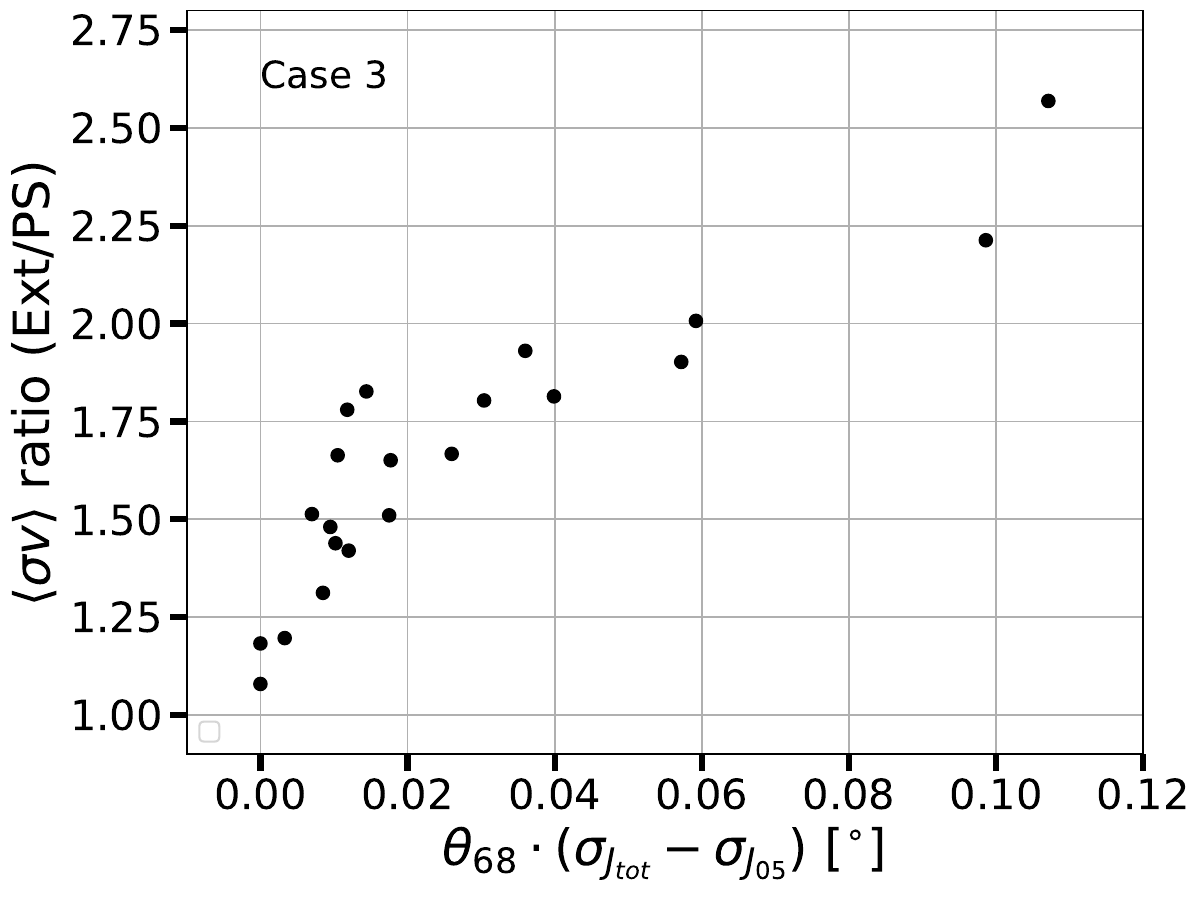}
\caption{Ratio of \sv~limits for the \texttt{Ext}/\texttt{PS} cases as a function of the parameter $\theta_{68}\cdot(\sigma_{J_{tot}}-\sigma_{J_{05}})$ as reported in Tab.~\ref{tab:jfactors}. We show the results obtained for the {\tt Case 3}.}
\label{fig:correlation3}
\end{figure}

Finally in \texttt{Case 3} (bottom left), we consider the effect of extension and of the difference in the average and error of the geometrical factors. In this case the general behaviour is the same presented before for \texttt{Case 1} and \texttt{Case 2}. However, the ranking of the dSphs with the largest \sv~ratio between the \texttt{Ext} and \texttt{PS} scenarios is driven mainly by the objects for which the difference of $\sigma_{J}$ is the largest, i.e.~Reticulum II and Coma Berenices.
This is explained by the \Jf-factor likelihood term, Eq.~\ref{eq:Lsigma}, which disfavors values of $J_i$ much different from the observed one.
Objects with $\sigma_{J_{\rm{tot}}}$ much larger than $\sigma_{J_{\rm{05}}}$, such as Reticulum II, have a likelihood profile that is broader for the \texttt{Ext} with respect to \texttt{PS} scenarios. 
We can understand this by thinking that the likelihood profile for counts, derived with Poisson statistics, is multiplied by a term related to the geometrical factor see Eq.~\ref{eq:geom}. Therefore, the larger $\sigma_{J}$ is in that equation, the broader is the shape of the likelihood as a function of the annihilation cross section, assuming a fixed DM mass.
This makes the upper limits found for the former model larger than the one of the latter and, as a result, the ratio \texttt{Ext}/\texttt{PS} is significantly larger than 1.
In this case the stacked analysis gives values of the ratio that are at most around 2 for a DM mass of about 300 GeV.
In order to demonstrate how the results depend on both $\theta_{68}$ and $\sigma_J$, we show, in Fig.~\ref{fig:correlation3}, the ratio of \sv~limits for the \texttt{Ext}/\texttt{PS} cases as a function of the combination of parameters $\theta_{68}\cdot(\sigma_{J_{tot}}-\sigma_{J_{05}})$, as reported in Tab.~\ref{tab:jfactors}. A clear correlation between the upper limits and the quantity $\theta_{68}\cdot(\sigma_{J_{tot}}-\sigma_{J_{05}})$ is present.

It might seem surprising that for objects such as Reticulum II or Coma Berenices the change from $\sigma_{J_{\rm{05}}}$ to $\sigma_{J_{\rm{tot}}}$ is much more important than the change in the average observed $J$ factor. We expect the tidal radius $r_{\rm t}$ to be responsible for the change since it is the only parameter that contributes to $J_{\rm tot}$ and not to $J_{05}$.
Indeed, we find that the objects having the largest change in $\sigma_{J}$ present two common characteristics: They have a significant extension ($\theta_{68}$) \textit{and} the posterior PDF of the tidal radius $r_{\rm t}$ is broad. Since for the ultra-faint dSphs $r_{\rm t}$ is computed from the fitting parameters $\rho_{\rm s}$, $r_{\rm s}$ and $D$ (see Sec.~\ref{sec:model}), any reduction of the error on these parameters from more accurate data or new analyses would reduce the error on $r_{\rm t}$ and affect the results for \texttt{Case 3}.

Finally, in the bottom right panel of Fig.~\ref{fig:sim_ratio}, we show, for \texttt{Case 3} the \sv~ratio for the stacked case (red line),
together with the 95\% and 68\% C.L.~bands, as obtained for the simulated data.

While in this section we have discussed the effects obtained when isolating the parameter value which distinguish the \texttt{Ext} case from the \texttt{PS} one, we stress that the \texttt{Ext} scenario is fully identified by the self-consistent choice of (a) an extended spatial template, (b) the normalization of the \Jf~factor to  $J_{\rm tot}$, and (c) the corresponding error on the \Jf~factor $\sigma_{J_{\rm tot}}$.  In what follows, all results  therefore refer to the parameters' values choice as in \texttt{Case 3}.

\section{Results with real data}
\label{sec:resdsphs}

\subsection{Detection significance}
\label{sec:detection}

\begin{figure*}
\includegraphics[width=0.45\textwidth]{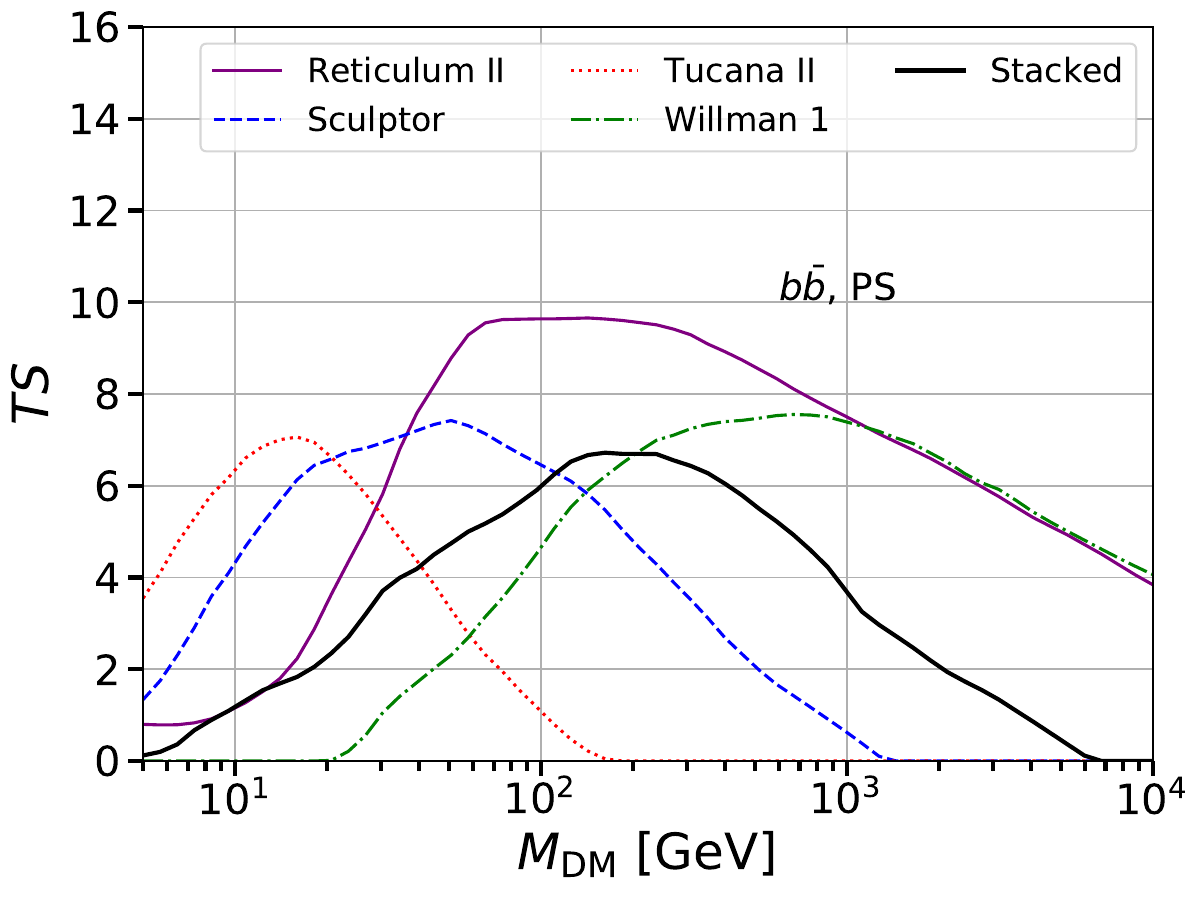}
\includegraphics[width=0.45\textwidth]{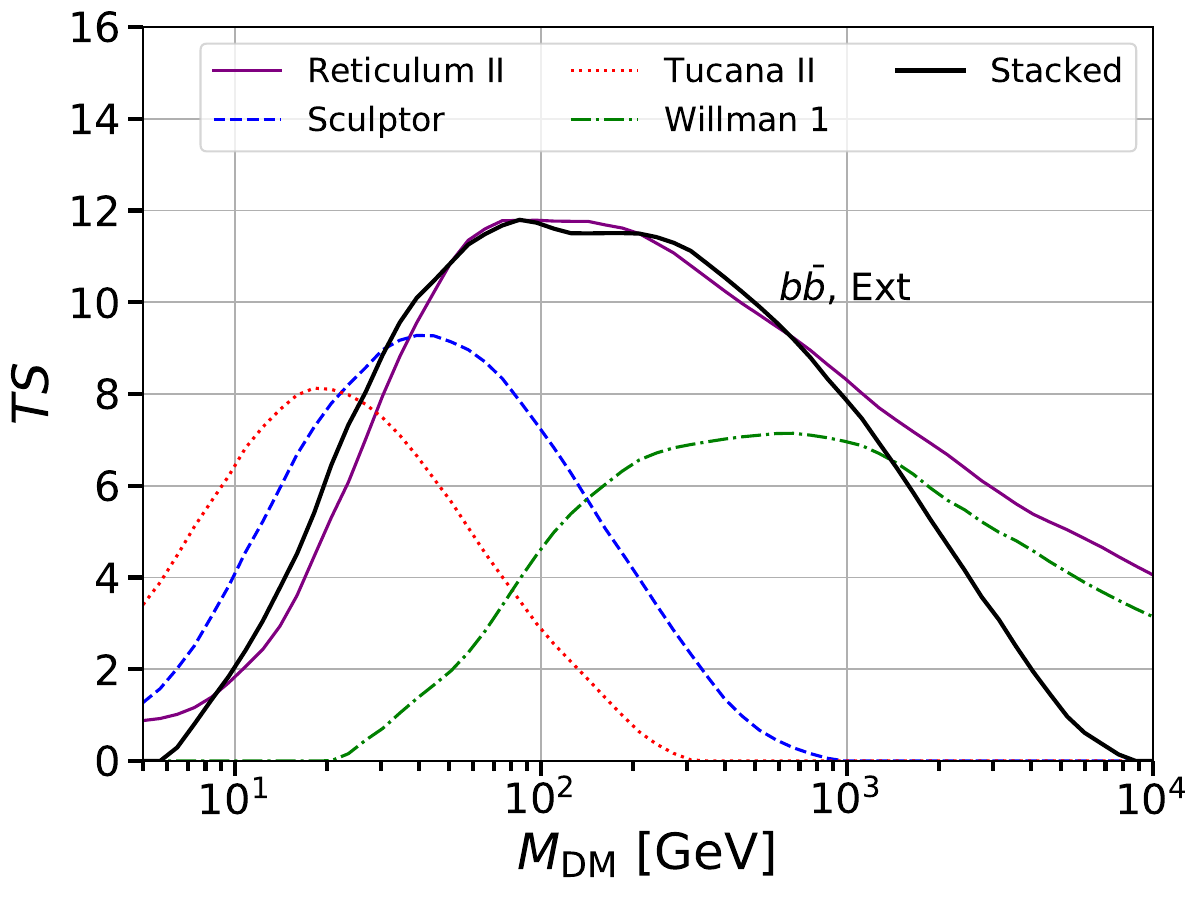}
\includegraphics[width=0.45\textwidth]{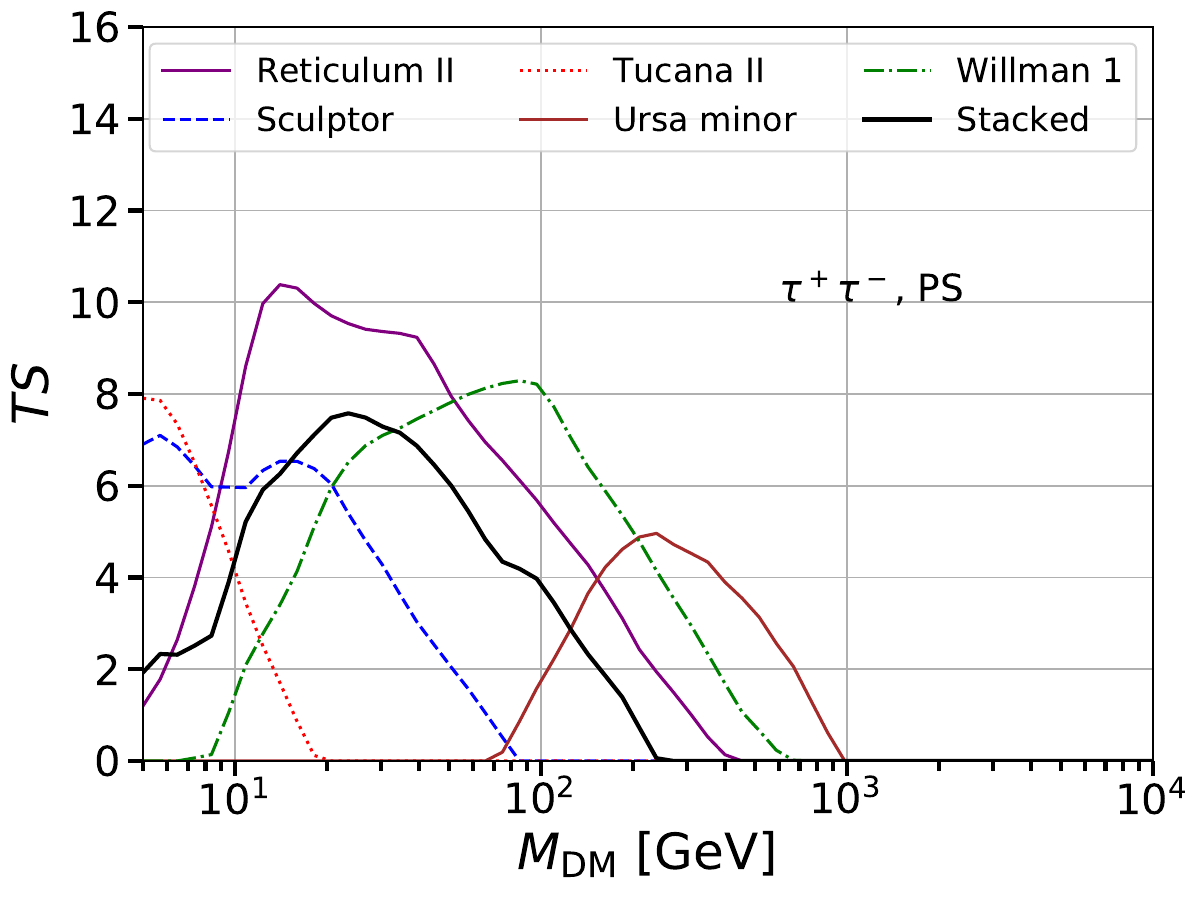}
\includegraphics[width=0.45\textwidth]{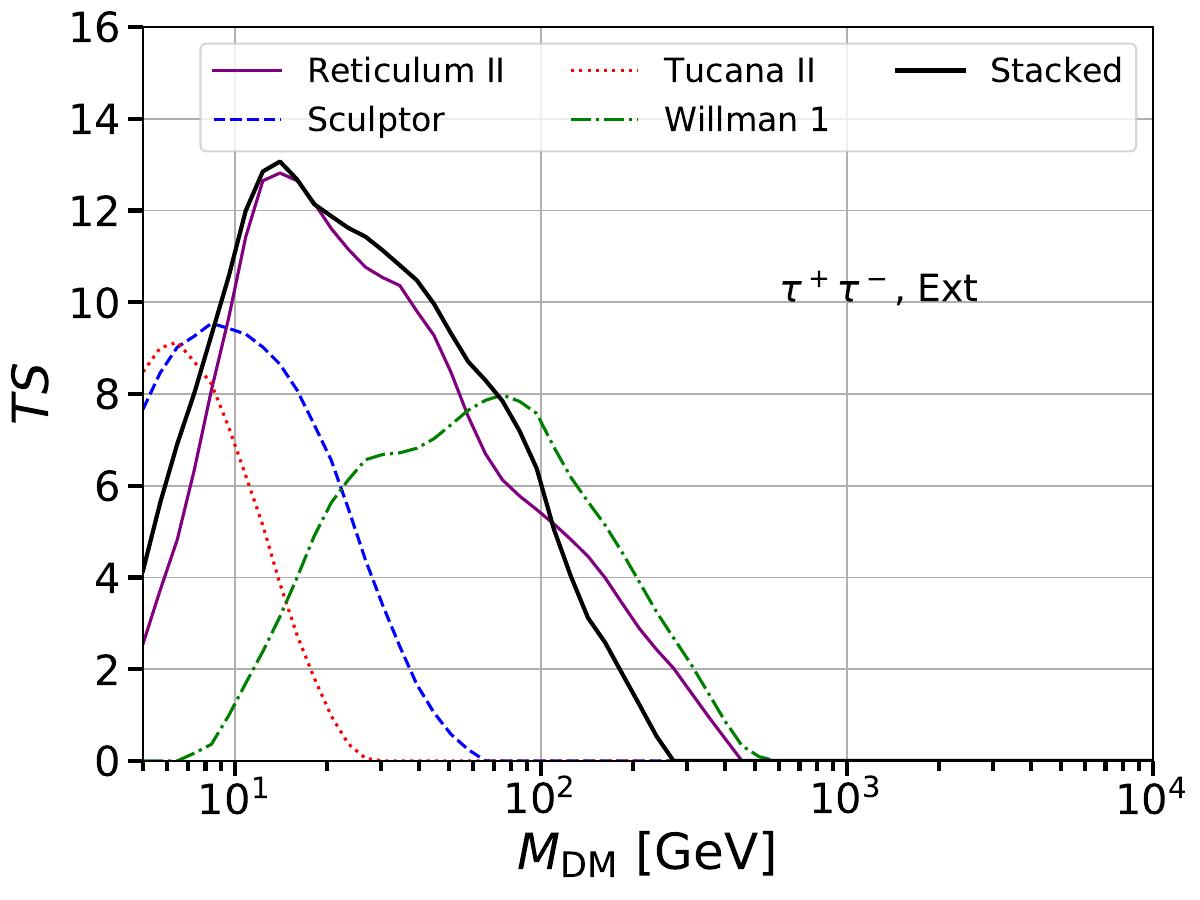}
\caption{{\it Real data}: Total $TS$ as a function of the DM mass for the dSphs detected with the highest significance. We show the results for $b\bar{b}$ (top panels) and $\tau^+\tau^-$ (bottom panels) annihilation channels and  for the \texttt{PS} (left panels) and \texttt{Ext} scenarios (right panels).}
\label{fig:TSdwarfs}
\end{figure*}

We first test the evidence of an additional source template (\texttt{PS} or \texttt{EXT}) at the position of each dSph in real data, see description in Sec.~\ref{sec:datanalysis}.
The $TS$ as a function of DM mass is displayed in Fig.~\ref{fig:TSdwarfs}
for the case of annihilation into $b$ quarks  (top panels) and $\tau$ leptons (bottom panels), and for the \texttt{PS} (left) and \texttt{Ext} (right) source model. 
We only show the dSphs detected with the highest significance, although this is never significant enough to claim evidence for an excess of photons -- the maximal, total, $TS$ reached is about 13, which roughly corresponds to $\sqrt{13} \sim 3.6 \, \sigma$ local significance (without considering degradation due to trial factors).
Among the dSphs selected, the one detected with the highest $TS$ in the \texttt{Ext} scenario is Reticulum II for a DM particle mass of of 50 -- 200 (10 -- 20) GeV, $\langle \sigma v \rangle = 1.3 \times 10^{-26}$ ($4\times 10^{-27}$) cm$^3$/s for the $b\bar{b}$ ($\tau^+\tau^-$) annihilation channel and detected with a $TS\sim 12$ (13), which corresponds to a p-value of $1.2 \times 10^{-3}$ ($ 7.5 \times 10^{-4}$) local, i.e.~pre-trials, significance of $\sim 3.0\sigma$ ($3.1\sigma$)
\footnote{In order to convert the $TS$ into the p-value and the detection significance, we have assumed that the $TS$ distribution of the null hypothesis is equal to the $\chi^2/2$ for 2 degrees of freedom, i.e.~the DM mass and annihilation cross section.}, in agreement with 
previous results, e.g.~\cite{Hooper:2015ula}. 

We also show in Fig.~\ref{fig:TSdwarfs} the $TS$ as a function of DM mass obtained with the combined analysis from all the dSphs in our sample.
In case of a real DM signal, we would observe a peak of the $TS$ which is higher than what was found from the individual sources.
We find that the maximum $TS$ we obtain, assuming an extended DM template for all dSphs in our sample, is 12 (13) for the $b\bar{b}$ ($\tau^+\tau^-$) annihilation channel in the \texttt{Ext} case. 
%
We find slightly smaller TS values for the \texttt{PS} case. This is consistent with the fact that the extended templates pick up more photons and residuals in the analysis and, as a result, the signal is found with a slightly larger significance.

\subsection{Upper limits on $\langle \sigma v \rangle$}
\label{sec:UL}

\begin{figure}
\includegraphics[width=0.45\textwidth]{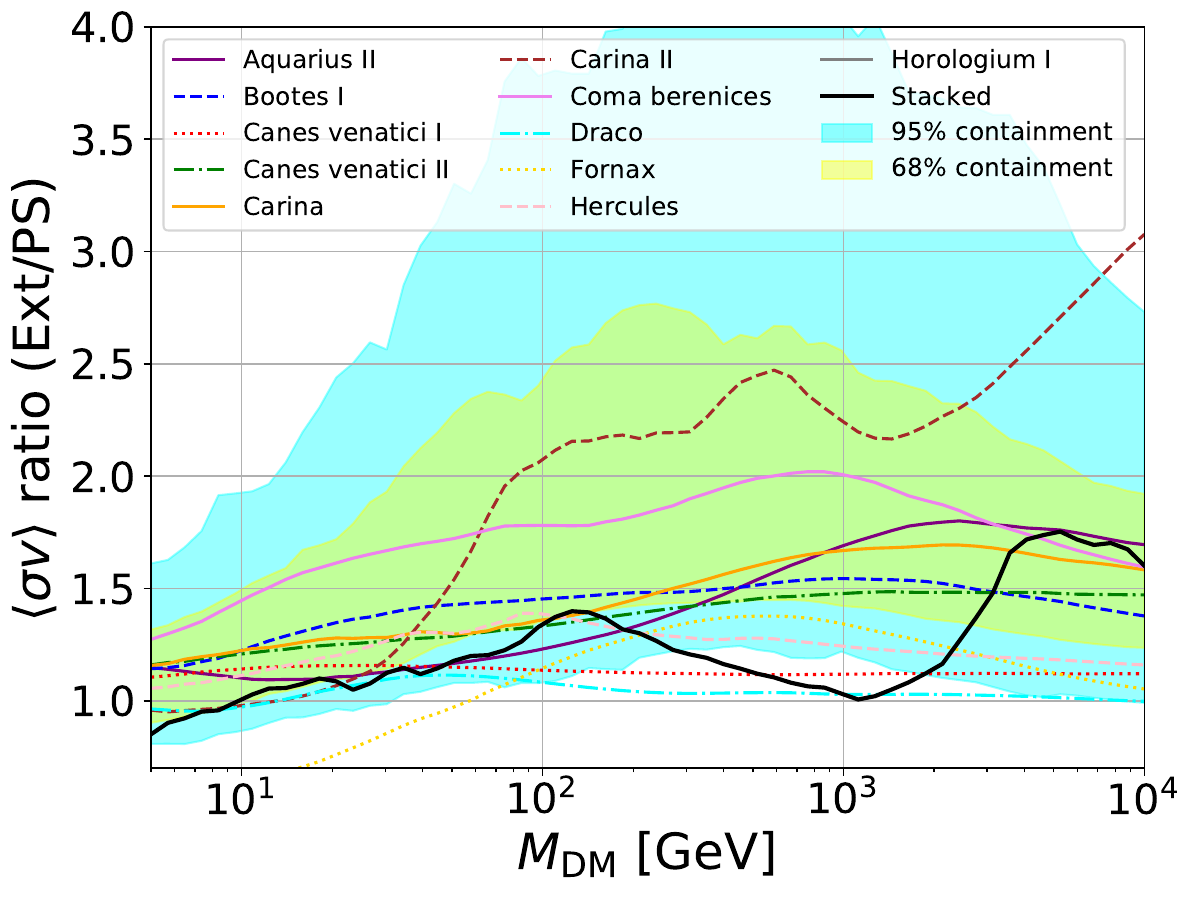}
\includegraphics[width=0.45\textwidth]{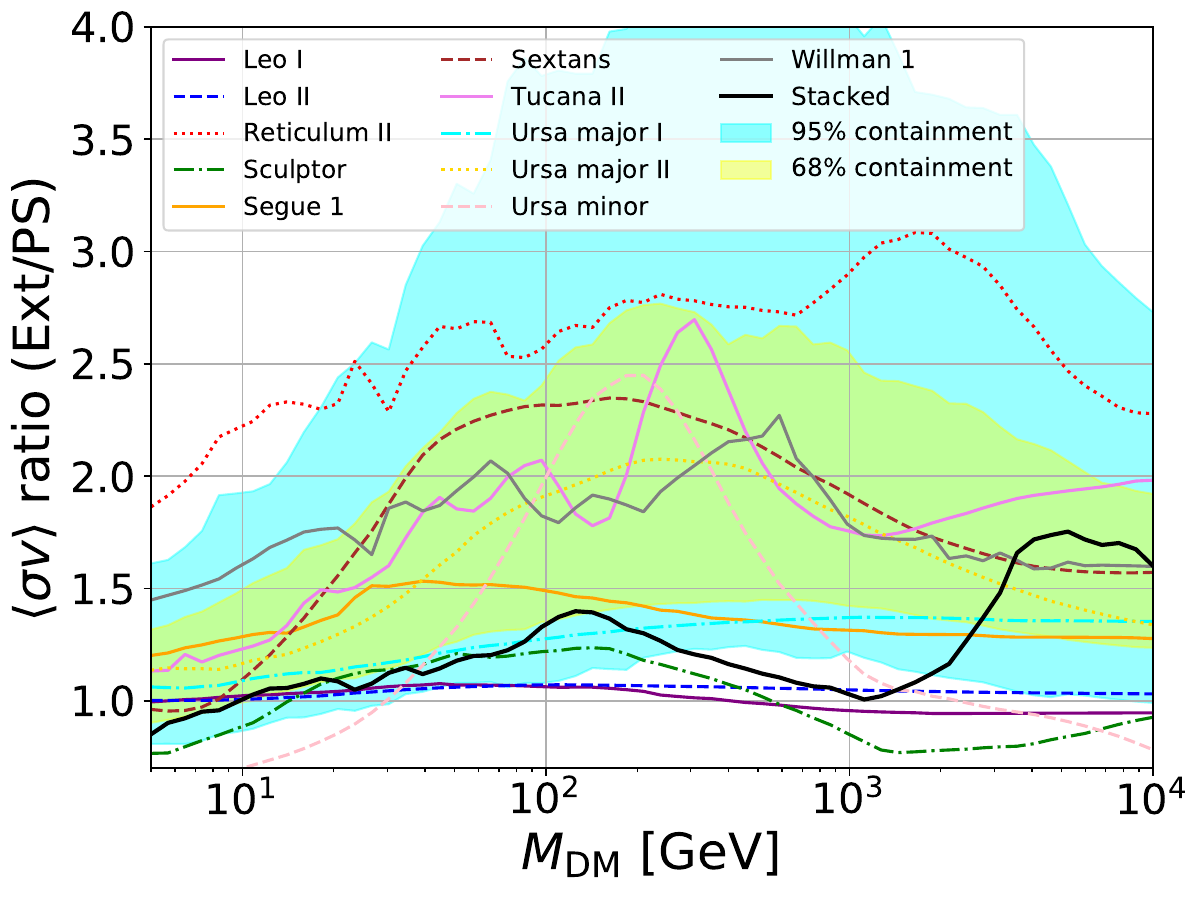}
\caption{{\it Real data}: Ratio between the $95\%$ C.L.~upper limits on \sv~found with the extended and point-like scenarios for the $b\bar{b}$ annihilation channel.
We show ratios for all individual dSphs, as well as for the stacked analysis (black line). 
The bands correspond to the 68\% -- 95\% C.L.~for the null hypothesis. The top and bottom panels report for legibility purpose two sets of dwarfs in our sample.}
\label{fig:sigmavratio}
\end{figure}

\begin{figure}
\includegraphics[width=0.49\textwidth]{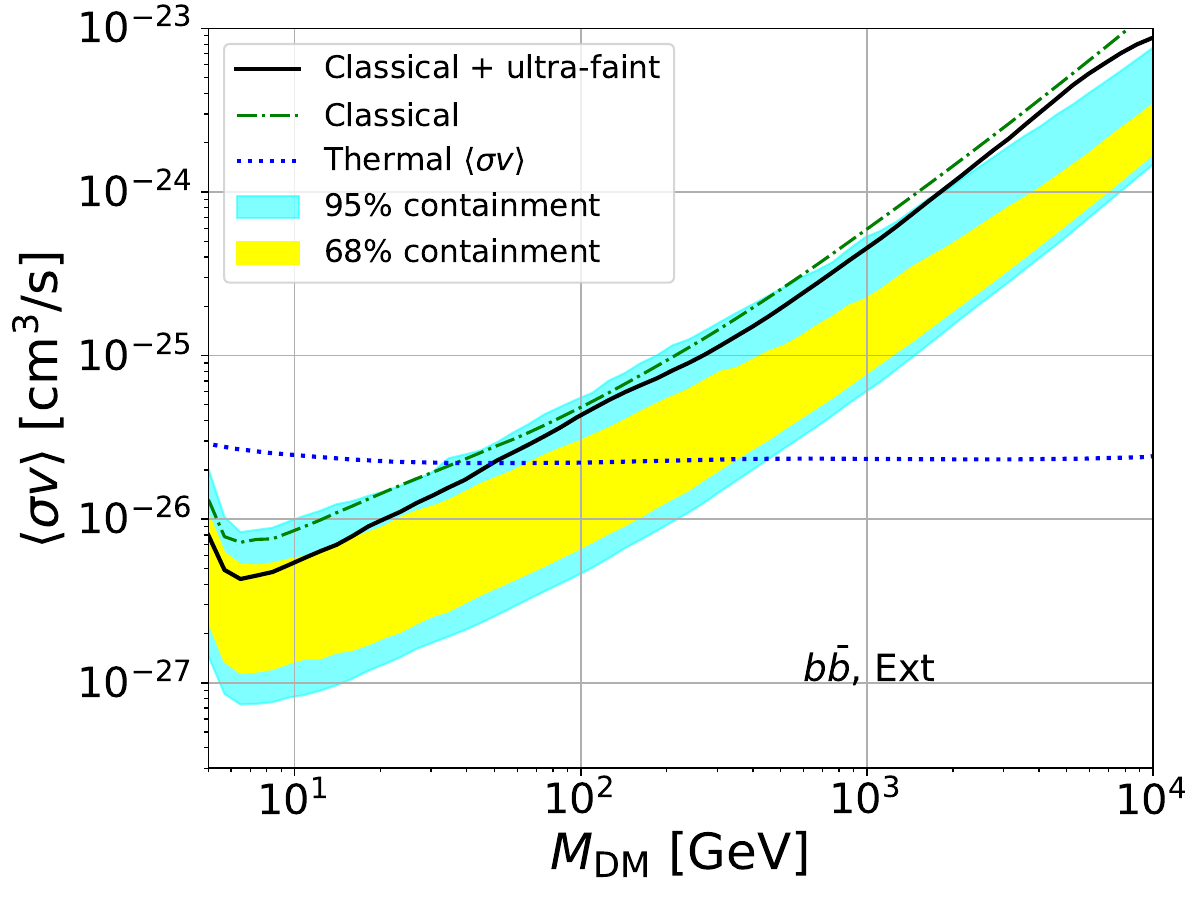}
\includegraphics[width=0.49\textwidth]{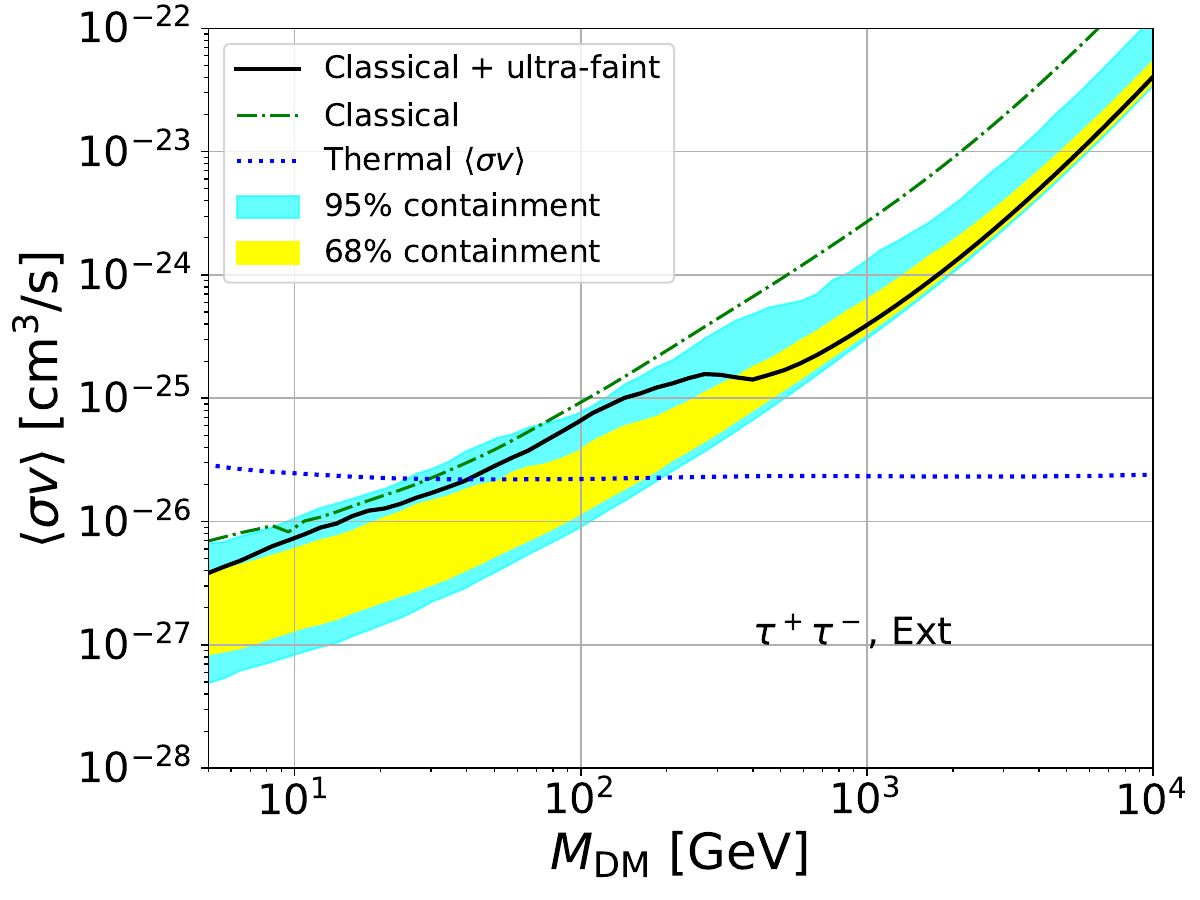}
\caption{{\it Real data}: 95\% C.L.~upper limits on the DM annihilation cross section, \sv, for annihilation into $b$ quarks (top panel) and $\tau$ leptons (bottom panel) in the \texttt{Ext} scenario. The stacked limit derived from the sample of 22 dSphs, classical and Ultra-faint (UF), is represented by the black solid line. The 68\% and 95\% C.L.~containment bands represent the distribution of the same limits under the null hypothesis. We also show the combined limit when only the 8 classical dSphs are considered (green dot-dashed line).  The thermal cross section is taken from~\cite{Steigman:2012nb} (blue dotted).
}
\label{fig:sigmadwarfs}
\end{figure}

\begin{figure}
\includegraphics[width=0.49\textwidth]{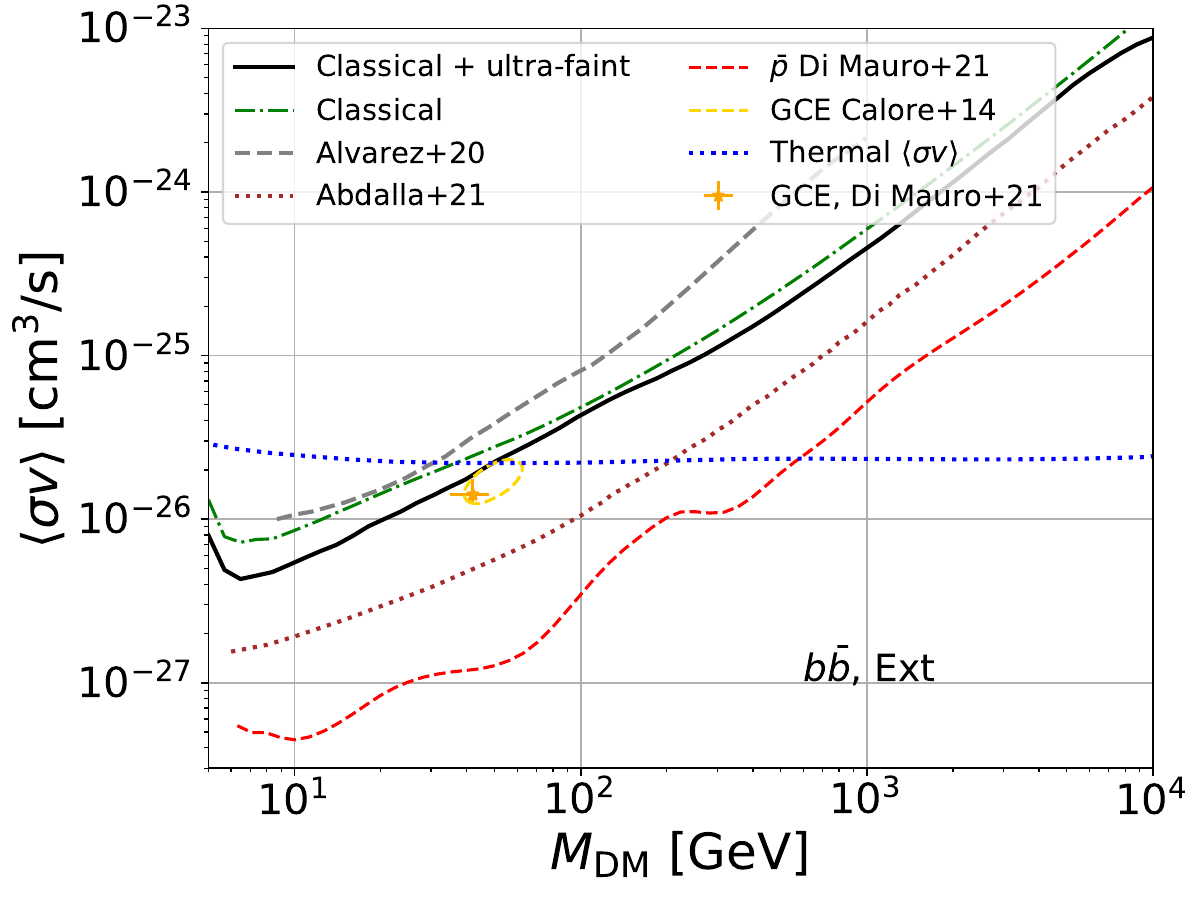}
\caption{95\% C.L.~upper limits on the DM annihilation cross section, \sv, for annihilation into $b$ quarks in the \texttt{Ext} scenario found with our analysis (black solid line). We also show the combined limit when only the 8 classical dSphs are considered (green dot-dashed line).  The thermal cross section is taken from~\cite{Steigman:2012nb} (blue dotted).
As a comparison we report the limits found for classical dSphs in \cite{Alvarez:2020cmw} (grey dashed) and the combined limits found for different gamma-ray experiments in \cite{Hess:2021cdp} (brown dotted).
We also overlay anti-proton limits~\cite{DiMauro:2021qcf}, and the best-fit 1$\sigma$ data point for the DM interpretation of the GeV excess taken from Refs~\cite{DiMauro:2021qcf,Calore:2014nla}.
}
\label{fig:sigmadwarfscomp}
\end{figure}

Since the signal detected from each individual dSph and for the stacked sample is not significant, we calculate upper limits for the 
annihilation cross section, \sv.
We do so for both the \texttt{PS} and \texttt{Ext} scenarios.
Analogously to what was done with simulated data, we show in Fig.~\ref{fig:sigmavratio}
the ratio of the limit on \sv~using an extended template over the one in the \texttt{PS} limit.
We assume DM particles annihilating into $b\bar{b}$ quarks.
We display the ratio for single dSphs and for the stacked case, together with the 68\% and 95\% C.L.~bands obtained with the simulations for the null signal.
The observed ratios for individual dSphs are mostly contained in the expectation bands. 
Nonetheless, there are cases where the ratio lies outside the bands.
For example the limit ratios found between a DM mass of 300-3000 GeV are slightly below the $95\%$ containment band.
This is also the case of Sculptor which, at masses of about 1 TeV, is below the $95\%$ containment band. 
We stress that the width of the bands here is only indicative and does not include possible effects such as background mismodeling. In fact, we remind that the simulations are performed with mock data assuming a perfect knowledge of the background sources and interstellar emission.
Therefore, the fact that some curves are above or below the bands could be due to a imperfect knowledge of the background components in the analysis of real data.
%

In general, with real data analysis, the ratios between the \sv~obtained with the \texttt{Ext} template and the one found with the \texttt{PS} case are closer to 1 than what we obtain with simulations, with ratios for single dSphs that reach at most 3 (1.6 for combined limits).
However, the result of the real data analysis is compatible with the $95\%$ C.L.~containment band derived from simulations.
In particular, for DM masses above 20 GeV the ratio between \texttt{Ext} and \texttt{PS} is a factor of about 1.7 smaller with respect to what we obtain for the average of the simulations.
The main reason for this result is that the \texttt{PS} case is less compatible with the null hypothesis results than the extended case. In other words, in real data the \texttt{PS} limits are weaker than in simulated data because the small signal detected for the point-like source case is in real data more significant with respect to the null hypothesis compared to what occurs in the extended scenario.
This implies that assuming an extended template for the DM emission makes the limits for $\langle \sigma v \rangle$ more compatible with the null detection.

Finally, we present the limits on \sv~as a function of DM mass in Fig.~\ref{fig:sigmadwarfs} for the $b\bar{b}$ (top panel) and $\tau^+\tau^-$ (bottom panel) annihilation channels for the \texttt{Ext} scenario.
We stress that this is the source model that better matches the 
characteristics of simulated dSphs, and this is therefore the model one has
to adopt in order to provide robust and self-consistent constraints from dSphs.
The stacked limit derived from the sample of 22 dSphs is represented by the black solid line. The 68\% and 95\% C.L.~containment bands represent the distribution of the limits under the null hypothesis.
The upper limits obtained are systematically higher than the $95\%$ containment band obtained with the simulations for $M_{\rm{DM}}>25$ GeV for the $b\bar{b}$ channel, and between 10 -- 300 GeV for the $\tau^+\tau^-$ channel.
The reason for this is related to the presence of small excesses as shown in Fig.~\ref{fig:TSdwarfs}. 
The $95\%$ C.L.~upper limits are below the thermal cross section~\cite{Steigman:2012nb} up to roughly 10 GeV for both channels.
Our results for the upper limits with dSphs are similar at the $20-30\%$ level with recently published in Refs~\cite{Calore:2018sdx,HoofEtAl2020,Alvarez:2020cmw,DiMauro:2021qcf} where different list of sources and analysis techniques have been applied. 
For a more direct comparison, we also show the combined limit when only the 8 classical dSphs are considered (green dot-dashed line).
We notice that our limits are comparable with~\cite{Alvarez:2020cmw} for the classical sample
although we do not perform a profiling over background uncertainties which can 
nonetheless impact the limits up to a factor of 3 for high masses (see Fig.~\ref{fig:sigmadwarfscomp}).
Instead, the limits reported recently in Ref.~\cite{Hess:2021cdp} from a combined analysis of {\it Fermi}-LAT, HESS, VERITAS, HAWC and MAGIC data look a factor of about 3 more stringent than ours. This is mainly due to the choice of the geometrical factor values and their uncertainties, and the sample of dSphs considered that differs from ours.
We also show, in Fig.~\ref{fig:sigmadwarfscomp}, the comparison of the upper limits found in this paper compared with the best-fit region for the DM parameters that fit the Galactic center excess well. We see that the upper limits we find are only slightly above the values of $\langle \sigma v \rangle$ that are compatible with the Galactic center excess. This demonstrates the importance of properly including the extension in the DM template for dSphs to correctly interpret this excess.

\subsection{Systematic uncertainties from non-spherical templates}
\label{sec:sys_asymm}

We report here the results obtained using the triaxial template introduced in Sec.~\ref{sec:triaxial_template}. The analysis is performed for Ursa Minor, which is one of the dSphs most impacted by the use of an extended template in place of a point-like one. We recall that three specific orientations are considered for the dSph, with the l.o.s.~being aligned with either the major, second or minor axes. The values of the different axes are $a = 1.28$ (major axis), $b = 1.02$ (second axis) and $c  = 0.78$ (minor axis). These values for $a$, $b$ and $c$ satisfy at a few $\%$ level the conditions between $b/a$, $c/a$ and $abc$ reported in Sec.~\ref{sec:triaxial_template}.

In the first configuration, the halo is less extended because the axes perpendicular to the l.o.s. are the second and minor ones. We also have ${\rm log}_{10}(J_{05}) = 18.36$ (in GeV$^2$/cm$^5$) which is 12\% higher than the spherical value ${\rm log}_{10}(J_{05}) = 18.31$. We recall that the values of the profile parameters (e.g.~$\rho_{\rm s}$ and $r_{\rm s}$) are the same for the spherical and the triaxial templates. 
In the second configuration, the major and minor axes are perpendicular to the l.o.s., while ${\rm log}_{10}(J_{05}) = 18.3$ is very close to the spherical value.
Finally, in the third configuration, the halo is more extended and ${\rm log}_{10}(J_{05}) = 18.23$ which is 20\% lower than the spherical case. A similar dependence of the $J$-factor on the orientation of the halo and comparable quantitative variations are found in the triaxial analyses of Refs.~\cite{BonnivardEtAl2015a,SandersEtAl2016}.

\begin{figure}
\includegraphics[width=0.49\textwidth]{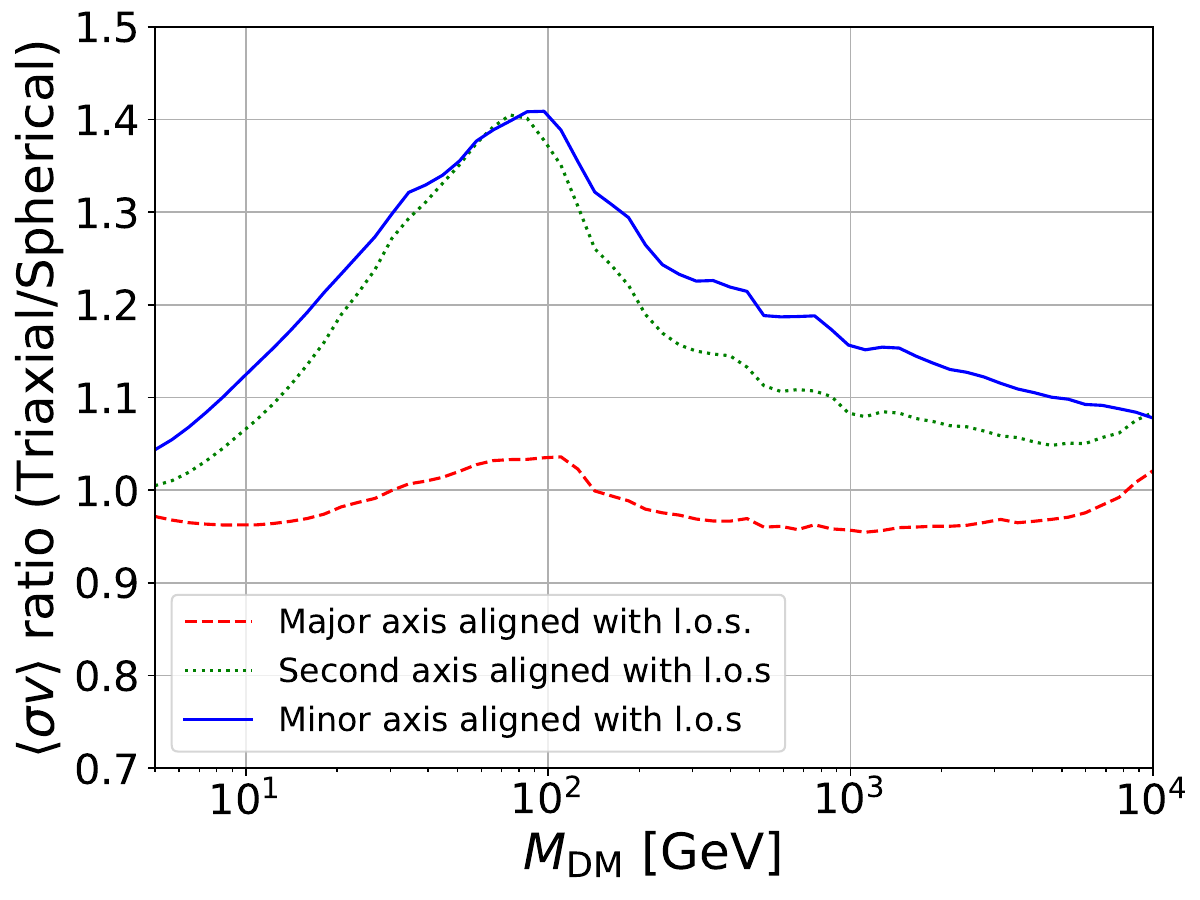}
\caption{Ratio between the 95\% C.L. upper limits on $\left<\sigma v\right>$ found with the extended triaxial and extended spherical scenarios for the $b\overline{b}$ annihilation channel for three different orientations of the Ursa Minor DM halo, see text for more details.}
\label{fig:triaxial_ratio}
\end{figure}

Ratios between the cross-section exclusion limits obtained with the spherical template and the triaxial one are shown in Fig.~\ref{fig:triaxial_ratio}. 
The case where the major axis is aligned with the l.o.s.~is represented by the dashed-red curve while the second- and minor-axis alignment cases are displayed by the dotted-green curve and blue curve, respectively.
We notice that the configuration where the major axis is oriented along the l.o.s. leads to a limit that is very similar to the spherical one (within $5\%$), while the second axis and minor axis orientations lead to cross-section upper limits that are higher by almost $40\%$ at a DM mass of 100 GeV.
This shows that the spatial morphology of the signal impacts the limit, not just the $J$-factor.
If the $J$-factor alone was the only relevant parameter, the \sv~ratio for major-axis orientation would be smaller than 1, the second-axis orientation would lead to a ratio very close to 1, and the ratio for the minor-axis orientation would be higher. 
This hierarchy between the different orientations is indeed observed in Fig.~\ref{fig:triaxial_ratio} but the \sv~ratio is shifted upward compared to expectations based on the $J$-factor alone. The ratio is also not flat, and peaks at 100 GeV. 
The triaxial template thus leads to constraints that are comparable or slightly weaker than the spherical ones. One should keep in mind that the orientations considered here correspond to extreme configurations as there is no reason why one of the main axes should be aligned with the l.o.s. for any given target, thus a 40$\%$ weakening of the limit should be seen as a maximal effect of triaxiality. 
We stress again that our analysis assumes that the DM halo structural parameters are the same in the spherical and triaxial case. 
A non-spherical Jeans analysis on the same kinematic data would probably lead to different values for these parameters, which would lead to different $J$-factors. We have shown however that the $J$-factor is not the only source of change and that morphology also plays a role.

\section{Conclusions}
\label{sec:conclusions}
According to the predictions of numerical and semi-analytical simulations, dSphs as the most massive DM sub-halos must have a sizeable angular extension.
In turn, the gamma-ray signal from DM annihilation in these objects is expected not to be point-like, as typically assumed in the literature, but rather extended in the sky.

In the present work, we first quantify what is the angular extension of a large sample of dSphs using the latest models of the DM distribution in these objects.
We found that 8 out of 22 dSphs have an effective angular size larger than the nominal \Fermi-LAT angular resolution at a few GeV, 
which motivated testing the impact of the adoption of an extended spatial template with a thorough gamma-ray data analysis.\footnote{The nominal sensitivity of the LAT at GeV energies taken from \url{https://www.slac.stanford.edu/exp/glast/groups/canda/lat_Performance.htm} is about 0.5 deg} and 8 dSphs in our sample have $\theta_{68}$ larger than $0.5^{\circ}$.

The extension, as defined here, is an effective parameter which is ultimately related to
the distance and the DM dSphs profile. For the same distance, the cuspier the profile is, the smaller the angular extension will be. However, we stress that we rely on state-of-art determination of the 
dSphs mass modeling and DM profile.

We defined a fully self-consistent model for, what we called, the \texttt{Ext} (extended) scenario, that is spherically symmetric, and we quantified the impact of using such a source model against
the traditionally adopted point-like source model, when looking for
excess of photons from the dSphs directions. 
We demonstrated that accounting properly for the dSphs angular extension has a significant impact on the limits on the DM annihilation cross section.
When considering the combined analysis of 
22 dSphs, for DM masses larger than 10 -- 15 GeV, the limits weaken by a factor up to 1.5 in the extended case, while for low masses
limits with an extended template are compatible with (or slightly 
stronger than) for the point-like case.
The mass dependence of the ratio can be understood as follows: the {\it Fermi}-LAT PSF is much larger at low energies than at high energy. On the other hand, the peak of the gamma-ray flux from DM moves at higher energies when the DM mass increases. Therefore, low-mass DM models are detected with a poorer PSF with respect to high-mass candidates making the \texttt{Ext} and \texttt{PS} more similar for low mass values.
For the individual dSphs analysis, instead, variations up to a factor of 3 less are induced by adopting an extended model for the dSph emission.

Such an effect is similar, in size, to other uncertainties that 
have been demonstrated in the past to affect (weaken) the robustness of the dSphs gamma-ray constraints, either related to 
the DM distribution in these objects~\cite{BonnivardEtAl2015a,UllioEtAl2016,SandersEtAl2016,BonnivardEtAl2016,HayashiEtAl2016,KlopEtAl2016,IchikawaEtAl2017}, or to the (mis-)modeling of the
astrophysical background at the dSph position~\cite{Mazziotta:2012ux,GeringerSamethEtAl2015,BoddyEtAl2018,Calore:2018sdx,Alvarez:2020cmw}.

We also test our analysis with a triaxial DM model. We find that the orientation of the axis could weaken the limits by at most a factor of 30-40$\%$ at around 100 GeV.

More generally, our limits are competitive with the ones from other targets such as the the Milky Way halo \cite{2012ApJ...761...91A,Huang:2015rlu,Zechlin:2017uzo,Chang:2018bpt} and the Galactic center (see, e.g., \cite{TheFermi-LAT:2017vmf,DiMauro:2021qcf}), while constraints from other messengers such as anti-protons (see, e.g., \cite{DiMauro:2021qcf,Calore:2022stf}) and from radio wavelengths~\cite{Regis:2021glv} keep 
setting the strongest limits on WIMP DM, even if they are typically more subject to astrophysical uncertainties such as the ones related to the cosmic-ray propagation or to the strength of magnetic field.

Our constraints, as it is for other limits
from gamma-ray searches towards dSphs, are only mildly in tension with the DM interpretation of the \Fermi~GeV excess detected towards the Galactic center, see e.g.~\cite{Calore:2014nla, DiMauro:2021qcf}. This tension can be alleviated when considering, among others, uncertainties on the Galactic DM halo distribution~\cite{Benito:2016kyp, Benito:2019ngh}.

In conclusion, we stress that spatial extension is a common feature of close-by, massive satellites, as shown in~\cite{DiMauro:2020uos}, and we recommend the community to take this effect into account when deriving limits from such objects with high-energy photons. 
As we have shown here, the impact of extension is relevant for dSphs. Compared to dSphs, we expect the impact on dark sub-halos to be 
less important, because of the correlation between \Jf-factor and extension, but still present. Ref.~\cite{DiMauro:2020uos} assessed the impact of extension on dark sub-halo detection, but we expect an impact also on the limits on DM particle models set through searches for  sub-haloes in unidentified \Fermi~sources.
Finally, we comment that galaxy clusters are also good targets for DM detection and should be rather extended, see discussion in~\cite{2012JCAP...07..017A, 2022arXiv220316440L}. 
In the end, the extended analysis of DM targets is undoubtedly of relevance of \Fermi-LAT searches, and will be even more so for the next generation gamma-ray telescope, i.e.~the Cherenkov Telescope Array, CTA \cite{CTAConsortium:2017dvg}.


\medskip

\begin{acknowledgments}
We warmly thank J.~Read for providing the MCMC chains 
from Ref.~\cite{Alvarez:2020cmw}, and A.~Pace for helpful discussions regarding Ref.~\cite{2019MNRAS.482.3480P}.
MDM research is supported by Fellini - Fellowship for Innovation at INFN, funded by the European Union’s Horizon 2020 research programme under the Marie Skłodowska-Curie Cofund Action, grant agreement no.~754496.
We warmly thank Viviana Gammaldi for careful reading of the manuscript and helpful comments.
FC and MS acknowledge support by the Programme National Hautes \'Energies (PNHE) through the AO INSU 2019, grant ``DMSubG" (PI: F. Calore). 
Visits of MDM to LAPTh were supported by Université Savoie Mont-Blanc, grant ``DISE" (PI: F. Calore).

The {\it Fermi} LAT Collaboration acknowledges generous ongoing support from a number of agencies and institutes that have supported both the development and the operation of the LAT as well as scientific data analysis. These include the National Aeronautics and Space Administration and the Department of Energy in the United States, the Commissariat\'a l'Energie Atomique and the Centre National de la Recherche Scientifique / Institut National de Physique Nucl\'eaire et de Physique des Particules in France, the Agenzia Spaziale Italiana and the Istituto Nazionale di Fisica Nucleare in Italy, the Ministry of Education, Culture, Sports, Science and Technology (MEXT), High Energy Accelerator Research Organization (KEK) and Japan Aerospace Exploration Agency (JAXA) in Japan, and the K. A. Wallenberg Foundation, the Swedish Research Council and the Swedish National Space Board in Sweden.
Additional support for science analysis during the operations phase is gratefully acknowledged from the Istituto Nazionale di Astrofisica in Italy and the Centre National d'Etudes Spatiales in France. This work performed in part under DOE Contract DE- AC02-76SF00515.
\end{acknowledgments}


\appendix

\section{Density profile parameters}
\label{app:profile_parameters}

\begin{table*}[]
    \centering
    \begin{tabular}{l|c|c|c|c|c|c}
           & ${\rm log_{10}}(\rho_{\rm s})$ & $r_{\rm s}$ & $r_{\rm c}$ & $n$ & $\delta$ & $r_{\rm t}$ \\
          & [$\rm M_\odot/kpc^3$] & [$\rm kpc$]  & [$\rm kpc$] & &  & [$\rm kpc$] \\  
 \hline
 Ursa Minor & 7.305 & 2.169 & 1.398 & 0.7852 & 4.289 & 1.113\\
 Draco & 7.341 & 1.678 & 0.1943 & 0.5389 & 4.159 & 1.114\\
 Sculptor & 7.360 & 2.136 & 1.370 & 0.7730 & 4.261 & 1.534 \\
 Sextans & 7.408 & 1.134 & 0.5343 & 0.5857 & 4.218 & 1.257\\
 Leo I & 7.355 & 1.483 & 0.4084 & 0.5209 & 4.257 & 1.374 \\
 Leo II & 7.614 & 0.9740 & 0.2283 & 0.5083 & 4.212 & 0.5436 \\
 Carina & 7.160 & 1.655 & 0.5988 & 0.5263 & 4.215 & 1.609 \\
 Fornax & 7.071 & 2.750 & 1.940 & 0.8600 & 4.395 & 2.272 \\
 \hline
 Aquarius II & 7.546 & 1.007 & - & - & - & 10.41\\
 Bootes I & 7.003 & 1.721 & - & - & - & 6.212 \\
 Canes Ven.\ I & 7.016 & 1.752 & - & - & - & 21.56 \\
 Canes Ven.\ II & 7.068 & 2.024 & - & - & - & 19.61 \\
 Carina II & 7.321 & 0.7256 & - & - & - & 2.097 \\
 Coma Beren. & 7.457 & 1.239 & - & - & - & 4.818 \\
 Hercules & 7.387 & 0.6921 & - & - & - & 7.340 \\
 Horologium I & 8.029 & 0.5642 & - & - & - & 7.459 \\
 Reticulum II & 7.545 & 0.8264 & - & - & - & 2.621 \\
 Segue 1 & 8.302 & 0.1921 & - & - & - & 1.139 \\
 Tucana II & 7.313 & 1.648 & - & - & - & 6.741  \\
 Ursa Major I & 7.425 & 1.115 & - & - & - & 9.910 \\
 Ursa Major II & 7.614 & 1.250 & - & - & - & 5.291 \\
 Willman 1 & 8.251 & 0.4534 & - & - & - & 3.736
    \end{tabular}
\caption{Sample of dSphs used in this study with the median value of their associated density profile parameters. DSphs in the top rows are taken from \cite{Alvarez:2020cmw}, {\it classical} dSphs, while dSphs in the bottom rows are taken from Ref.~\cite{2019MNRAS.482.3480P}, {\it ultra-faint} dSphs.
    }
    \label{tab:profile_parameters}
\end{table*}

\bibliography{main}

\end{document}